\documentclass[onefignum,onetabnum]{siamart220329}

\usepackage{lipsum}
\usepackage{amsfonts}
\usepackage{graphicx}
\usepackage{epstopdf}
\usepackage{algorithmic}
\ifpdf
  \DeclareGraphicsExtensions{.eps,.pdf,.png,.jpg}
\else
  \DeclareGraphicsExtensions{.eps}
\fi

\newsiamremark{remark}{Remark}
\newsiamremark{hypothesis}{Hypothesis}
\crefname{hypothesis}{Hypothesis}{Hypotheses}
\newsiamthm{claim}{Claim}

\headers{A Macroscopically Consistent Reactive Langevin Dynamics Model}{S. A. Isaacson, Q. Liu, K. Spiliopoulos, and C. Yao}

\title{A Macroscopically Consistent Reactive Langevin Dynamics Model\thanks{\funding{This work was partially supported by ARO W911NF-20-1-0244, W911NF2510078 and National Science Foundation DMS-2325185, DMS-2311500, and DMS-1902854.}}}

\author{Samuel A. Isaacson\thanks{Department of Mathematics and Statistics, Boston University, Boston, MA
(\email{isaacsas@bu.edu}, \email{liuq19@bu.edu}, \email{kspiliop@bu.edu}, \email{c2yao@bu.edu}).}
  \and Qianhan Liu\footnotemark[3]
  \and Konstantinos Spiliopoulos\footnotemark[3]
  \and Chen Yao\footnotemark[3]
  }

\usepackage{amsopn}

\ifpdf
\hypersetup{
  colorlinks=true,
  linkcolor=blue,
  citecolor=blue,
  pdftitle={A Macroscopically Consistent Reactive Langevin Dynamics Model},
  pdfauthor={S. A. Isaacson, Q. Liu, K. Spiliopoulos, and C. Yao}
}
\fi

\usepackage{amsmath, amssymb, amsfonts, mathtools, bm, bbm}
\usepackage{cite}
\usepackage{makecell}
\usepackage{changepage}
\usepackage{threeparttable} 
\usepackage{algorithmic}
\Crefname{ALC@unique}{Line}{Lines}
\usepackage{bookmark}

\newsiamremark{example}{Example}
\crefname{example}{Example}{Example}

\newsiamthm{assumption}{Assumption}
\crefname{assumption}{Assumption}{Assumption}

\usepackage{bbm}

\newcommand{\Lp}{\mathcal{L}}

\renewcommand{\vec}[1]{\boldsymbol{#1}}
\newcommand{\vecs}[1]{\boldsymbol{#1}}
\newcommand{\paren}[1]{\left(#1\right)}
\newcommand{\brac}[1]{\left[#1\right]}

\newcommand{\PD}[2]{\frac{\partial#1}{\partial#2}}

\newcommand{\lap}[1]{\Delta#1}

\newcommand{\abs}[1]{\left|#1\right|}

\newcommand{\kb}{k_{\text{B}}}

\newcommand{\veta}{\vecs{\eta}}

\newcommand{\vn}{\vec{n}}

\newcommand{\vP}{\vec{P}}

\newcommand{\va}{\vec{a}}
\newcommand{\vb}{\vec{b}}

\newcommand{\vx}{\vec{x}}

\newcommand{\vzeta}{\vec{\zeta}}

\newcommand{\vv}{\vec{v}}

\newcommand{\vxi}{\vec{\xi}}

\newcommand{\ind}{\mathbbm{1}}

\def\N{\mathbb{N}}
\def\R{\mathbb{R}}

\newcommand{\Kd}{K_{\textrm{d}}}

\renewcommand{\epsilon}{\varepsilon}

\newcommand{\GaussianDensity}{\mathcal{G}}

\newcommand{\fab}[1]{f_{1 2}^{(#1)}}
\newcommand{\fc}[1]{f_{3}^{(#1)}}

\newcommand{\pS}{\textrm{S}}
\newcommand{\sA}{\textrm{A}}
\newcommand{\sB}{\textrm{B}}
\newcommand{\sC}{\textrm{C}}
\newcommand{\sD}{\textrm{D}}

\newcommand{\Rpp}{\mathcal{R}_+}
\newcommand{\Rmm}{\mathcal{R}_-}
\newcommand{\Ipp}{\mathbb{I}_+}
\newcommand{\Imm}{\mathbb{I}_-}

\newcommand{\vnp}{\vec{n}^+}
\newcommand{\vnm}{\vec{n}^-}
\newcommand{\vN}{\mathbf{N}}

\newcommand{\ina}{\vec{i}^{\vn}_{\va}}
\newcommand{\inb}{\vec{i}^{\vn}_{\vb}}
\newcommand{\xina}{\vec{\xi}^{\vn}_{\va}}
\newcommand{\xinb}{\vec{\xi}^{\vn}_{\vb}}

\newcommand{\zetana}{\vec{\zeta}^{\vn}_{\va}}
\newcommand{\zetanb}{\vec{\zeta}^{\vn}_{\vb}}

\newcommand{\Xip}{\Xi_+}
\newcommand{\Xim}{\Xi_-}

\usepackage{xcolor}
\newcommand{\rev}[1]{\begingroup#1\endgroup}

\date{\today}
\begin{document}

\maketitle

\begin{abstract}
Particle-based stochastic reaction-diffusion (PBSRD) models are a popular
approach for capturing stochasticity in reaction and transport processes across
biological systems. In some contexts, the overdamped approximation inherent in
such models may be inappropriate, necessitating the use of more microscopic
Langevin Dynamics models for spatial transport. In this work we develop a novel
particle-based Reactive Langevin Dynamics (RLD) model, with a focus on
deriving reactive interaction kernels that are consistent with the physical
constraint of detailed balance of reactive fluxes at equilibrium. We demonstrate
that, to leading order, the overdamped limit of the resulting RLD model
corresponds to the volume reactivity PBSRD model, of which the well-known Doi
model is a particular instance. Our work provides a step towards systematically
deriving PBSRD models from more microscopic reaction models, and suggests
possible constraints on the latter to ensure consistency between the two
physical scales.
\end{abstract}



\section{Introduction}\label{s1}

The macroscopic, population-level dynamics of systems across cell, synthetic, and systems biology often arises from the stochastic movements of large collections of discrete entities or agents with short-range interactions \cite{Chakraborty2010kb, Nadkarni2012short, Naylor2017simbiotics, Siokis2018f, Sturrock2013spatial, WoldeEgfrdPNAS2010}. One popular framework to depict such dynamics are particle-based stochastic reaction-diffusion (PBSRD) models \cite{Chakraborty2010kb, Nadkarni2012short, Siokis2018f, Sturrock2013spatial}. PBSRD models are appropriate for studying chemical systems in cells containing millions of particles, over timescales of minutes to days. These models provide an intermediate framework between more microscopic quantum mechanical or molecular dynamics models, which are typically limited in scale and computationally intensive \cite{ShawAntonMS2009}, and more macroscopic mean-field chemical kinetics models described by deterministic reaction-diffusion PDEs. Volume reactivity (VR) models as popularized by Doi \cite{DoiSecondQuantA, DoiSecondQuantB, TeramotoDoiModel1967} are a commonly used PBSRD model. They model the movements of particles by Brownian Dynamics, and particle interactions by reactive interaction kernels, which encode the probability density per time that a reaction occurs based on the current positions of forward and backward substrates.

Though VR PBSRD models provide an effective description for stochastic
reaction-diffusion systems, their reliance on overdamped Brownian dynamics,
which assumes inertial effects are negligible, limits their applicability in
numerous realistic settings. \rev{This assumption fails in systems requiring
fine temporal resolution, containing particles with large mass, or involving
low-friction environments where velocity correlations and memory effects play a
critical role. In such contexts, ignoring inertia can lead to both qualitatively
and quantitatively inaccurate results \cite{lowen2020inertial}. Instead,
Langevin dynamics (LDs), which explicitly incorporates particle velocities and
inertial forces, becomes indispensable for systems where motion is underdamped,
such as macroscopic self-propelled particles (including viborobots and granulates
\cite{scholz2018inertial}) or mesoscopic particles in low-viscosity media like
gases (e.g., dust particles in plasmas \cite{morfill2009complex}).

Beyond modeling particle transport, integrating LDs with particle reactions is
needed for accurately modeling the growth of bacterial
populations~\cite{Naylor2017simbiotics}, reactive-transport processes in
cellular systems where reaction timescales are comparable to the timescale for
relaxation of inertial forces~\cite{HolmesCerfonPRL2022}, and the dynamics of
insect populations that can undergo swarming and
flocking~\cite{Ariel2015locust}. For instance, \textit{Simbiotics}
\cite{Naylor2017simbiotics} is a platform for 3D modeling of bacterial
populations, which models bacterial motion using LDs while accounting for
interactions of bacteria with the environment and other cells. In this context,
accounting for velocity and inertial effects is important due to the variety of
mechanical, chemical, and viscous forces experienced by cells. In such systems,
reactions are a natural way to model random phenotypic changes in cells, or
cell-cell interactions that can change cell states (such as one cell killing
another or undergoing division). More generally, reaction models are useful for
any type of population process in which individual particles/agents can change
state, and/or interact to change state (including interactions that result in
the creation or removal of some particles). For example, in insect models such
interactions can be used to model reproduction, death, or predator-prey
interactions.}

While PBSRD models have been extensively studied and validated against
experimental data \cite{Huhn2023molecular} and more macroscopic theories
\cite{ErbanChapman2009, Hanggi1990reaction, Lipkova2011analysis}, the literature
on how to represent reactions in LD-scale models is more limited
\cite{Burschka1981kinetic, IsaacsonErban2016, Kneller1985boundary}. \rev{To help
bridge this gap, in this work we focus on developing a simple particle-based
reactive Langevin dynamics (RLD) model, with the goal of constructing this model
such that it is consistent with accepted VR PBSRD models in the overdamped
limit. Our work represents a step towards constructing physically-consistent
reaction models for use in more complicated Langevin Dynamics-based models, such
as used in modeling cell populations~\cite{Naylor2017simbiotics}, or as used in
developing coarse-grained particle approximations to more detailed molecular
dynamics models that are needed to more accurately resolve the spatial reaction
dynamics of cellular processes~\cite{erban2020coarse}.}


The core of developing RLD models is then in constructing reactive interaction kernels for which solutions to the RLD model converge in the overdamped limit to solutions of the VR PBSRD model with standard (overdamped) reactive interaction kernels. The desired RLD kernels can be decomposed into two components: (a) reactive rate functions, representing the probability per time that substrates will react based on their current positions, and (b) placement densities, which represent the probability density that reaction products are placed at specific locations with specific velocities, given the locations and velocities of substrates.

The contribution of this work is two-fold. First, assuming conservation of momentum and pointwise detailed balance of reaction fluxes at equilibrium for reversible reactions, we derive concrete, novel formulas for reactive interaction kernels in general reactive Langevin dynamics models. For the reader's convenience, we have summarized the formulas we derive for three common reversible systems in \cref{s1 tab: rate functions}-\cref{s1 tab: v placement}. Second, using these kernels, we derive the (high-friction/small-mass) overdamped limit via asymptotic expansions of solutions to the RLD model, and show that the leading order terms satisfy the equations of the VR PBSRD model. This establishes that our RLD models are consistent with VR PBSRD models in the overdamped limit. While we propose a particular family of reactive interaction kernels in this work, for example assuming conservation of momentum during reactive collisions, the scalings we obtain also suggest how alternative kernels could be constructed that still maintain consistency with standard overdamped VR PBSRD models.

The paper is organized as follows: In \cref{s2}, we establish the basic setting
for RLD models in a multi-particle system for general mass action reactions.
We present motivating examples to introduce the new reactive rate functions and
placement densities. In \cref{s3}, we construct the forward Kolmogorov equation
governing the evolution of the probability density for the system to be in a
given state, we derive the general reversible reaction detailed balance
condition at equilibrium, we illustrate how detailed balance constrains
reversible reaction interaction kernels, and we state our assumptions on the
reactive interaction kernels for general systems.  In \cref{s4}, we derive the
overdamped limit of RLD models by developing asymptotic expansions of the
solution to the forward equation in the limit of large damping constant. We
demonstrate that to leading order, the asymptotic expansion of the marginal
density that projects out the velocity component satisfies the standard forward
equation for the overdamped VR PBSRD model. In \cref{s5}, we demonstrate how our
theory translates in the case of the common reversible reactions  $\sA+\sB
\rightleftharpoons \sC$ and $\sA+\sB \rightleftharpoons \sC+\sD$. In particular,
we derive the detailed-balance consistent forward and backward reactive
interaction kernels presented in \cref{s2}. We also sketch how the overdampled
limit of the RLD model in each of these special cases recovers the VR PBSRD
model, giving a less notationally-heavy sketch of the more general calculation
of ~\cref{s4}.
To validate our theoretical results, numerical simulations are carried out for
the $\sA+\sB \rightleftharpoons \sC$ reaction in \cref{s6}. Conclusions and
pointers to future work are included in \cref{s7}.

\section{Notation and motivation}\label{s2}

Let us consider a system of $J$ biochemical species, labeled by $\pS_1, ..., \pS_J$,
with $N_j(t)$ denoting the stochastic process for the number of particles of species $j$
at time $t$, and $\vN(t) = \big(N_1(t),...,N_J(t)\big)$ the population state vector for
all species. We denote $n_j$ as a value for $N_j(t)$ and $\vn = (n_1,...,n_J)$
as a value for $\vN(t)$. Denote the
positions and velocities of $n_j$ particles of $\pS_j$ at time $t$ by
\begin{equation*}
	\rev{\vec{X}}^{(j)}(t) = \big(X^{(j)}_1(t), ..., X^{(j)}_{n_j}(t)\big) \in \R^{n_j d},
	\qquad
	\rev{\vec{V}}^{(j)}(t) = \big(V^{(j)}_1(t), ..., V^{(j)}_{n_j}(t)\big) \in \R^{n_j d}.
\end{equation*}

Each particle moves within a domain $\Omega \subset \R^{d}$
according to the Langevin equations
\begin{equation}\label{s2 eq: Langevin Dynamics}
	\dot{X}^{(j)}_l(t) = V^{(j)}_l(t),
	\qquad
	\dot{V}^{(j)}_l(t) = -\beta_j V^{(j)}_l(t) + \beta_j \sqrt{2D_j} \dot{W}^{(j)}_l(t),
\end{equation}
where each $W^{(j)}_l$ is a standard Brownian Motion, $\beta_j$ is the scaled friction
constant of species-$j$ with ``per time" units, and $D_j$ is the diffusion coefficient constant
of species $j$. We further assume that these constants are related via Einstein's relation
\begin{equation}\label{s2 eq: Einsten relation}
	m_j D_j \beta_j = k_B T,
\end{equation}
where $m_j$ denotes the mass of particles of type $j$, $k_B$ is the Boltzmann
constant and $T$ is a fixed constant representing temperature. In what follows,
unless stated otherwise, we will assume that $\Omega$ is finite. \rev{We will
also assume that at the boundary a reflecting Neumann boundary condition holds,
but expect that our general results should also apply for periodic boundary
conditions (which we use in our subsequent numerical illustrations). Note that
our asymptotic expansion to derive the general over-damped limit in \cref{s3}
focuses on and holds for the interior of the domain, see the conclusion
\cref{s7} for more discussion on the impact of boundary conditions.}

Possible values for the stochastic processes $X^{(j)}(t)$ and $V^{(j)}(t)$ are denoted by
\begin{equation*}
	\vx^{n_j} = \left(x^{(j)}_1, ..., x^{(j)}_{n_j}\right) \in \Omega^{n_j},
	\qquad
	\vv^{n_j} = \left(v^{(j)}_1, ..., v^{(j)}_{n_j}\right) \in \R^{n_j d}.
\end{equation*}
We define the state of a particle by the collective position and velocity pair, labeled by
$\xi^{(j)}_l := (x^{(j)}_l, v^{(j)}_l)$. The collection of states of all
particles given the population state vector, $\vn$, is then denoted by
$\vxi^{\vn} = (\vxi^{n_1}, ..., \vxi^{n_J})$. Similarly, we can define
$\vx^{\vn}$ and $\vv^{\vn}$ as the collection of positions and velocities of all
particles. Assume particles of the same species are indistinguishable, i.e., a
state $\widetilde{\vxi}^{\vn}$ is equivalent to $\vxi^{\vn}$ if, for each specie
$j$, the state vector $\widetilde{\vxi}^{n_j}$ is simply a reordering of
$\vxi^{n_j}$.

For any given population state $\vn$, we let $p^{\vn}(\vxi^{\vn}, t)$
denote the probability density that $\vN(t)=\vn$ with the particles located at some state
equivalent to $\vxi^{\vn}$. Hence,
\begin{equation*}
	\mathbb{P}(\vN(t)=\vn) = \frac{1}{\vn !} \int_{(\Omega \times \R^d)^{|\vn|}} p^{\vn}(\vxi^{\vn}, t) \mathrm{d} \vxi^{\vn},
\end{equation*}
where the factorial, \rev{$\vn ! := n_1 ! \, n_2! \cdots n_J ! $}, arises from
overcounting indistinguishable particle states.
Finally, we let $\vP(t) = \{p^{\vn}(\vxi^{\vn}, t)\}_{\vn}$ represent
the vector of probability densities over all possible states at time $t$.

In addition to the spatial motion of each particle governed by scaled Langevin dynamics,
\rev{we also allow arbitrary order reversible reactions between particles}
\begin{equation}\label{s2 eq: reaction}
	a_1 \pS_1 + a_2 \pS_2 + \cdots + a_J \pS_J
	\rightleftharpoons
	b_1 \pS_1 + b_2 \pS_2 + \cdots + b_J \pS_J ,
\end{equation}
where $\va = (a_1,...,a_J)$ labels the \rev{forward substrate} stoichiometry vector, $\vb
= (b_1,...,b_J)$ labels the \rev{backward substrate} stoichiometry vector, and
\rev{$|\va| = a_1 + a_2 +... a_J$ is the order of forward reaction (with $|\vb|$ defined analogously for the backward reaction).}
As we will frequently encounter Maxwell-Boltzmann (i.e. Gaussian) distributions, to
ease notation and make explicit the Gaussian nature of the distribution, we shall denote the corresponding probability density (with zero mean) by
\begin{align}
 \GaussianDensity_{q}(x;\sigma^{2}\mathrm{I}_{q})&=\frac{1}{(2\pi\sigma^{2})^{q/2}}e^{-\frac{\abs{x}^{2}}{2\sigma^{2}}}, \text{ for }x\in\mathbb{R}^{q},\label{Eq:GaussianDistpdf}
\end{align}
where  $\sigma^{2}\mathrm{I}_{q}$ is the variance-covariance matrix and $q$ is the dimension.

To illustrate the setting and introduce the notion of reactive interaction functions, we next present some  specific examples.

\begin{example}[$\sA + \sB \rightleftharpoons \sC$]\label{s2 eg: ABC}
	Consider a system consisting of \rev{different} species $\sA$, $\sB$ and $\sC$, which can undergo the reversible
	reaction $\sA+\sB \rightleftharpoons \sC$. For the forward reaction $\sA+\sB \rightarrow \sC$,
	denote the forward reaction rate function by $K^{\beta}_{+}(\xi_1, \xi_2)$, representing the
	probability per time that an $\sA$ particle at $\xi_1$ binds with a $\sB$ particle
	at $\xi_2$. Here, the $\beta$ superscript indicates a possible dependency on the friction
	constant. Analogously, for the reverse reaction $\sA+\sB \leftarrow \sC$, we can define
	backward rate function $K^{\beta}_{-}(\xi_3)$ representing the probability  per time
	that a $\sC$ particle at $\xi_3$ unbinds.
	
	To determine the positions and velocities of product particle $\rm{C}$ of reaction
	$\rm{A} + \rm{B} \rightarrow \rm{C}$, let $m^{\beta}_{+} (\xi_3|\xi_1, \xi_2)$
	denote the forward placement density that a product $\sC$ particle is placed at $\xi_3$
	given the substrates' states $\xi_1$ and $\xi_2$. $m^{\beta}_{+} (\xi_3|\xi_1, \xi_2)$ is assumed to be
	normalized so that
	\begin{equation*}
		\int_{\Omega \times \R^d} m^{\beta}_{+}(\xi_3|\xi_1, \xi_2) \, \mathrm{d} \xi_3 = 1.
	\end{equation*}
	The backward placement density $m^{\beta}_{-} (\xi_1, \xi_2|\xi_3)$ is defined analogously.
	We further assume the placement densities can be decomposed into a
	product as follows
	\begin{align*}
		m^{\beta}_{+} (\xi_3|\xi_1, \xi_2)
		&=
		m_{+} (x_3|x_1, x_2)
		m^{\beta}_{+} (v_3|v_1, v_2), \\
		m^{\beta}_{-} (\xi_1, \xi_2|\xi_3)
		&=
		m_{-} (x_1, x_2|x_3)
		m^{\beta}_{-} (v_1, v_2|v_3),
	\end{align*}
	where $m_{+} (x_3|x_1, x_2)$ and $m_{-} (x_1, x_2|x_3)$ are placement
	densities for positions, and $m^{\beta}_{+} (v_3|v_1, v_2)$ and
	$m^{\beta}_{-} (v_1, v_2|v_3)$ are placement densities for velocities. In
	the remainder, each of these placement densities are assumed to be
	properly normalized.
	\begin{remark}\label{s2 remark: m} For the sake of brevity, we use the same
		notation $m_{+}^{\beta}(\cdot \mid \cdot)$ to represent the probability
		density of the first argument given the second argument, regardless of
		whether these arguments pertain to position, velocity, or state. \rev{
		Additionally, we assume that the spatial placement densities for
		particle positions are independent of the friction constant $\beta$, and
		as such we do not write the $\beta$-superscript for them. The lack of
		$\beta$-dependence of the spatial placement densities is illustrated in
		the specific choices of the following examples. In contrast, the
		velocity placement densities will typically depend on $\beta$,
		hence we maintain the $\beta$-superscript for them.  }
	\end{remark}

	A common choice for $K^{\beta}_{-}(\xi_3)$ in the overdamped case would be a constant rate, i.e.,
	$K^{\beta}_{-}(\xi_3) := K_{-}(x_3) = \lambda_{-}$.
	To define the forward rate function, a common model is that
	the two particles bind with some constant rate, $\lambda_+$, when their distance falls within
	a specified reaction radius $\varepsilon>0$, i.e. the Doi model \cite{DoiSecondQuantA, DoiSecondQuantB}
	\begin{equation}\label{s2 eq: AB-C K plus}
		K^{\beta}_{+} (\xi_1, \xi_2) := K_{+} (x_1, x_2) = \lambda_{+} \ind_{[0, \varepsilon]}(|x_1 - x_2|).
	\end{equation}
	Note, both rate functions depend solely on positions, and are independent of the
	friction constant $\beta$.

	For the forward position placement density, the product $\sC$ is chosen to
	lie at some point along the line segment connecting $\sA$ and $\sB$, i.e.
	\begin{equation}\label{s2 eq: AB-C m x plus}
		m_{+} (x_3|x_1, x_2) = \delta (x_3 - (\alpha x_1 + (1-\alpha) x_2)),
	\end{equation}
	where \rev{$\delta(\cdot)$ is the Dirac delta function} and $\alpha \in [0, 1]$ is fixed.
	One common choice for $\alpha$ is the diffusion weighted center of mass, $D_2 / (D_1 + D_2)$,
	see~\cite{ZhangIsaacson2022}. For the forward velocity placement, we assume
	conservation of momentum holds, and hence we have
	\begin{equation}\label{s2 eq: AB-C m v plus}
		m^{\beta}_{+} (v_3|v_1, v_2) = \delta \Big(v_3 - \tfrac{m_1 v_1 + m_2 v_2}{m_3} \Big),
	\end{equation}
	where $m_1$, $m_2$ and $m_3$ are masses of particles $\sA$, $\sB$ and $\sC$ respectively.

	For the backward $\sC \rightarrow \sA + \sB$ reaction, specifying the center
	of mass for the products via \cref{s2 eq: AB-C m x plus} is insufficient to
	uniquely determine their positions. We therefore also require that
	their separation, $x_1 - x_2$, is uniformly distributed within $B_\varepsilon$, the ball of
	radius $\varepsilon$ (with volume $\abs{B_{\varepsilon}}$). Hence, $m_- (x_1, x_2 |
	x_3)$ has the following form
	\begin{equation}\label{s2 eq: AB-C m x minus}
		m_- (x_1, x_2 | x_3)
		=
		\tfrac{1}{\abs{B_\varepsilon}}
		\ind_{[0,\varepsilon]} (\abs{x_1-x_2})
		\delta(x_3 - (\alpha x_1 + (1-\alpha) x_2)).
	\end{equation}
	Let $\Kd$ denote the dissociation constant for the reaction. As shown
	in~\cite{ZhangIsaacson2022}, using the preceding choice for $m_-$ and
	setting $\lambda_- := \Kd \lambda_+ \abs{B_{\varepsilon}}$ is consistent
	with detailed balance of pointwise reaction fluxes holding at equilibrium
	for the overdamped problem.

	Similarly, conservation of momentum is insufficient to uniquely specify the
	velocities of the $\sA$ and $\sB$ particles. We therefore derive one
	additional constraint from enforcing consistency with detailed balance of
	pointwise reaction fluxes holding at equilibrium, which we show in \cref{s3}
	and \cref{s5} gives that
	\begin{equation}\label{s2 eq: AB-C m v minus}
		m^{\beta}_{-} (v_1, v_2 | v_3)
		=
		\delta\paren{v_3 - \tfrac{(m_1 v_1 + m_2 v_2)}{m_3}}
        \GaussianDensity_{d}(v_1-v_2;(D_1\beta_1 + D_2\beta_2) \mathrm{I}_d).
	\end{equation}
	\rev{This model specifies the velocity distribution for products of reaction
	$\rm{C} \rightarrow \rm{A} + \rm{B}$ by enforcing conservation of total
	momentum via a Dirac delta constraint, and sampling the particles' velocity
	separation from an equilibrium Maxwell-Boltzmann distribution $\mathcal{N}
	(0, (D_1\beta_1 + D_2\beta_2) \mathrm{I}_d)$.}

	Finally, we note a useful scaling property of these specific velocity
	placement densities that we will later exploit in establishing the
	overdamped, i.e. $\beta \to \infty$, limit. Assume that $\beta_i = \beta \hat{\beta}_i$ and define
	$\gamma_i:=D_i\hat{\beta}_i$. By the Einstein Relation \cref{s2 eq: Einsten
	relation} and assuming conservation of mass, $m_1 + m_2 = m_3$, we have for i $\in
	\{1, 2\}$
	\begin{equation} \label{eq:a_b_to_c_d_m_gamma_idents}
		\frac{m_i}{m_3} = \frac{D_3 \beta_3}{D_i \beta_i} = \frac{\gamma_3}{\gamma_i},
		\quad \text{ and }\quad
       D_3 \beta_3
		= \frac{D_1 \beta_1 D_2 \beta_2}{D_1 \beta_1 + D_2 \beta_2}
		\Leftrightarrow
		\gamma_3
		= \frac{\gamma_1 \gamma_2}{\gamma_1 + \gamma_2}.
	\end{equation}

	Consider the change of variables, $v_i = \sqrt{\beta \gamma_i} \eta_i$ for
	$i = 1, 2, 3 $, representing a non-dimensional coordinate system in which we
	will study the over-damped limit in \cref{s4}. In these coordinates we have
	\begin{gather} \label{s2 eq: A-BC v place plus factor}
		\begin{aligned}
		m^{\beta}_{+} (v_3|v_1, v_2)
		&= \tfrac{1}{(\beta \gamma_3)^{d/2}}
		\delta \Big(\eta_3 - \Big(\sqrt{\tfrac{\gamma_3}{\gamma_1}} \eta_1 + \sqrt{\tfrac{\gamma_3}{\gamma_2}} \eta_2 \Big) \Big) \\
		&=: \tfrac{1}{(\beta \gamma_3)^{d/2}} \widetilde{m}_{+} (\eta_3|\eta_1, \eta_2),
		\end{aligned}\\
	 \label{s2 eq: A-BC v place minus factor}
		\begin{aligned}
			m_-^{\beta}(v_1, v_2 | v_3) &=
			\tfrac{1}{\beta^d (2\pi \gamma_1\gamma_2)^{d/2}}
			\delta \Big(\eta_3 - \Big(\sqrt{\tfrac{\gamma_3}{\gamma_1}} \eta_1 + \sqrt{\tfrac{\gamma_3}{\gamma_2}} \eta_2 \Big) \Big)
			e^{(\abs{\eta_3}^2-\abs{\eta_1}^2-\abs{\eta_2}^2)/2} \\
			&=: \tfrac{1}{\beta^d(\gamma_1 \gamma_2)^{d/2}} \widetilde{m}_{-} (\eta_1, \eta_2|\eta_3).
		\end{aligned}		
	\end{gather}
	The new coordinate system factors out the $\beta$ scaling from the
	$\beta$-dependent densities $m^{\beta}_{\pm}$. Consequently, the transformed
	densities $\widetilde{m}_{\pm}$ become independent of $\beta$. We will
	observe this scaling property for each of the specific reversible reactions
	we consider.
\end{example}

\begin{example}[$\sA + \sB \rightleftharpoons \sC + \sD$]\label{s2 eg: ABCD}
	Consider a system consisting of \rev{different} species $\sA$, $\sB$, $\sC$ and $\sD$, which
	can undergo the reversible reaction $\sA+\sB \rightleftharpoons \sC + \sD$.
	Similar to the previous example, we define the rate functions as
	\begin{align*}
		K^{\beta}_{+} (\xi_1, \xi_2) &:= K_{+} (x_1, x_2) = \lambda_{+} \ind_{[0, \varepsilon]}(|x_1 - x_2|)
			, \\
		K^{\beta}_{-} (\xi_3, \xi_4) &:= K_{-} (x_3, x_4) = \lambda_{-} \ind_{[0, \varepsilon]}(|x_3 - x_4|),
	\end{align*}
	where $\varepsilon >0$ represents the reaction radius.
	For the placement densities, we again assume the decomposition
	\begin{align*}
		m^{\beta}_{+} (\xi_3, \xi_4|\xi_1, \xi_2)
		&=
		m_{+} (x_3, x_4|x_1, x_2)
		m^{\beta}_{+} (v_3, v_4|v_1, v_2), \\
		m^{\beta}_{-} (\xi_1, \xi_2|\xi_3, \xi_4)
		&=
		m_{-} (x_1, x_2|x_3, x_4)
		m^{\beta}_{-} (v_1, v_2|v_3, v_4).
	\end{align*}

	Given two positions $x_1$ and $x_2$, the ordered pair $(x_3, x_4)$ coincides with
	positions pair $(x_1, x_2)$ or $(x_2, x_1)$ with the probability $p$ and $(1-p)$ respectively,
	that is
	\begin{equation} \label{eq:ABCDm_plus}
		m_+(x_3, x_4 | x_1, x_2)
		=  p \delta_{(x_1,x_2)}(x_3,x_4)+(1-p) \delta_{(x_1,x_2)}(x_4,x_3) ,
	\end{equation}
	\rev{where $\delta_{(x_1,x_2)}(x_3,x_4)$ denotes the bivariate Dirac delta, defined
	as $\delta_{(x_1,x_2)}(x_3,x_4)$ $:= \delta_{x_1}(x_3) \cdot \delta_{x_2}(x_4)$.}
	The backward position placement density $m_-(x_1, x_2 | x_3, x_4)$ is
	defined analogously by symmetry of the reaction. Let $\Kd$ now denote the
	dissociation constant for this reaction. By an analogous derivation to that
	in~\cite{ZhangIsaacson2022} for the preceding example, choosing $m_-$
	symmetrically to \cref{eq:ABCDm_plus} and setting $\lambda_- := \Kd
	\lambda_+$ is consistent with detailed balance of pointwise reaction fluxes
	holding at equilibrium for the overdamped problem.

	When considering the velocity placement of the forward reaction, the
	constraint of conservation of momentum is insufficient to uniquely specify
	the velocities $v_3$ and $v_4$ of the products $\sC$ and $\sD$ particles.
	Similar to \cref{s2 eg: ABC}, enforcing consistency with detailed balance
	fully determines the velocity placement density as (see \cref{s3} and
	\cref{s5})
	\begin{multline} \label{s2 eq: AB-CD m plus}
		m_+^{\beta}(v_3, v_4 | v_1, v_2)
		= (m_3 + m_4)^d \, \delta\paren{(m_3 v_3 + m_4 v_4)- (m_1 v_1 + m_2 v_2)}\\
        \times \GaussianDensity_{d}(v_3-v_4; (D_3\beta_3+D_4\beta_4)\mathrm{I}_{d}).
	\end{multline}
 	In this form, we can see that $m_+^{\beta}$ corresponds to placing the
    products such that total momentum is preserved, and their velocity
    separation is sampled from the Maxwell-Boltzmann distribution, i.e. $v_3 -
    v_4\sim \mathcal{N} (0, (D_3\beta_3 + D_4\beta_4) \mathrm{I}_d)$. The
    backward placement density $m_-^{\beta}(v_1, v_2 | v_3, v_4)$ can be defined
    analogously via the symmetry of the reaction. We show in \cref{s5} that
    this choice is consistent with pointwise detailed balance of the reaction
    fluxes holding at equilibrium.

	Finally, we demonstrate the $\beta$ scaling behavior of the velocity
	placement densities. Assume that $\beta_i = \beta \hat{\beta}_i$ and
	define $\gamma_i:=D_i\hat{\beta}_i$. The Einstein relation \cref{s2 eq:
	Einsten relation} can be rewritten as $m_i \gamma_i \beta = k_B T$, which
	implies $m_i / m_j = \gamma_j / \gamma_i$, for any $i \neq j$. Then,
	the Einstein relations and conservation of mass give that
	\begin{equation} \label{eq:a_b_to_c_d_param_ident_1}
	\begin{multlined}
		k_B T (m_1 + m_2) = D_1 \beta_1 m_1^{2}+ D_2 \beta_2 m_2^{2}
		= \frac{1}{\beta}(\kb T)^{2} \frac{\gamma_1+ \gamma_2}{\gamma_1 \gamma_2} \\
		= k_B T (m_3 + m_4) = D_3 \beta_3 m_3^{2}+ D_4 \beta_4 m_4^{2}=  \frac{1}{\beta}(\kb T)^{2} \frac{\gamma_3+ \gamma_4}{\gamma_3 \gamma_4}
	\end{multlined}
	\end{equation}
	and
	\begin{equation} \label{eq:a_b_to_c_d_param_ident_2}
 		\frac{D_3\beta_3 D_4\beta_4}{D_3\beta_3 + D_4\beta_4} = \frac{k_B T}{m_3 + m_4}
		= \frac{k_B T}{m_1 + m_2} = \frac{D_1\beta_1 D_2\beta_2}{D_1\beta_1 + D_2\beta_2}.
	\end{equation}

	Let $v_i = \sqrt{\beta \gamma_i} \eta_i$, for $i = 1,2,3,4$. After some
	algebra, we find that the velocity placement densities transform as
	\begin{align*}
		m_+^{\beta}(v_3, v_4 | v_1, v_2)
		&= \frac{1}{\beta^{d}} \left(\frac{\gamma_3+ \gamma_4}{2 \pi  \gamma_3^{2}  \gamma_4^{2}}\right)^{d/2}
		  e^{-\frac{\abs{\eta_3}^2}{ 2 }-\frac{\abs{\eta_4}^2}{ 2 }+\frac{\left| \frac{\eta_3}{\sqrt{\gamma_3}}+\frac{\eta_4}{\sqrt{\gamma_4}}\right|^{2}}{2\left(\frac{\gamma_3+ \gamma_4}{\gamma_3 \gamma_4}\right)}} \\
		  &\qquad\times \delta\paren{\left(\frac{\eta_3}{\sqrt{\gamma_3}}+\frac{\eta_4}{\sqrt{\gamma_4}}\right)- \left(\frac{\eta_1}{\sqrt{\gamma_1}}+\frac{\eta_2}{\sqrt{\gamma_2}}\right)}\nonumber\\
		&=: \frac{1}{\beta^d (\gamma_3\gamma_4)^{d/2}}\widetilde{m}_+(\eta_3, \eta_4 | \eta_1, \eta_2),
	\end{align*}
	where $\widetilde{m}_+(\eta_3, \eta_4 | \eta_1, \eta_2)$
	is of order one with respect to $\beta$. We can verify a similar scaling property holds for $m_-^{\beta}(v_1, v_2 | v_3, v_4)$.
\end{example}

Similar to the previous two examples, we can derive rate functions $K_{\pm}$ and
placement densities $m_{\pm}^{\beta}$ that are consistent with detailed balance
for the $\sA \rightleftharpoons \sB$ reversible reaction (again assuming
conservation of mass and momentum). In this case the velocity placement density
is just a $\delta$-function, which under the non-dimensional change of
coordinate $v_i = \sqrt{\beta \gamma_i} \eta_i, i=1,2$, scales as
\begin{equation*}
	m^{\beta}_{+}(v_2|v_1) = \delta(v_1 - v_2)
	= \frac{1}{(\beta \gamma_2)^{d/2}}
	\delta\Big(\eta_2 - \sqrt{\frac{\gamma_1}{\gamma_2}} \eta_1\Big)
	=: \frac{1}{(\beta \gamma_2)^{d/2}} \widetilde{m}_{+}(\eta_2 | \eta_1).
\end{equation*}

For all three of the preceding reactions, our concrete choices of rate functions
and placement densities are summarized in \cref{s1 tab: rate functions} through
\cref{s1 tab: v placement}. We emphasize that these choices are both consistent
with detailed balance of pointwise reaction fluxes holding at equilibrium, while
also maintaining consistency in the overdamped limit with common choices used in
PBSRD models (i.e. the rate functions of \cref{s1 tab: rate functions} and the
placement densities of \cref{s1 tab: x placement}). In addition to detailed
balance, they each arise from also assuming conservation of mass and momentum
during reactions.
\begin{table}[htbp]
    \caption{\rev{Reative Rate Functions} (with $\Kd$ the dissociation constant of the reaction)}
    \label{s1 tab: rate functions}
    \centering
    \begin{tabular}{| c | c | c |}
    \hline
    Reaction &
    $K_+(\vx^{\vn}_{\va})$ &
    $K_-(\vx^{\vn}_{\vb})$ \\ [0.3ex]
    \hline
    $\sA + \sB \rightleftharpoons \sC$ &
    $\lambda_{+} \ind_{[0, \varepsilon]}(|x_1 - x_2|)$ &
    $\lambda_- := \Kd \lambda_+ \abs{B_{\varepsilon}}$\\  [0.5ex]
    \hline
    $\sA \rightleftharpoons \sB$ &
    $\lambda_+$ &
    $\lambda_- := \Kd \lambda_+$ \\ [0.3ex]
    \hline
    $\sA + \sB \rightleftharpoons \sC + \sD$ &
    $\lambda_{+} \ind_{[0, \varepsilon]}(|x_1 - x_2|)$ &
    $\lambda_{-} \ind_{[0, \varepsilon]}(|x_3 - x_4|)$, $\lambda_- := \Kd \lambda_+$ \\  [0.3ex]
    \hline
    \end{tabular}

	\vspace{4pt}
    \footnotesize
    \rev{Note, $\vx^{\vn}_{\va}$ and $\vx^{\vn}_{\vb}$ denote general position
	vectors of the forward and backward reaction substrates. The detailed definition
	will be provided in the section \ref{s3-1}.}
\end{table}
\begin{table}[htbp]
	\centering
    \caption{Position Placement Densities}
    \label{s1 tab: x placement}
	\makebox[\linewidth]{
    \begin{tabular}{| c | c | c |}
    \hline
    Reaction &
    $m_+^{\beta}(\vx^{\vn}_{\vb} | \vx^{\vnp}_{\va})$ &
    $m_-^{\beta}(\vx^{\vn}_{\va} | \vx^{\vnm}_{\vb})$ \\ [0.5ex]
    \hline
    $\sA + \sB \rightleftharpoons \sC$ &
    $\delta (x_3 - (\alpha x_1 + (1-\alpha) x_2))$ &
    \makecell[l]{
        $\frac{1}{\abs{B_\varepsilon}} \ind_{[0,\varepsilon]} (\abs{x_1-x_2})$ \\
        $\, \times  \delta(x_3 - (\alpha x_1 + (1-\alpha) x_2))$
    }\\ [1.5ex]
    \hline
    $\sA \rightleftharpoons \sB$ &
    $\delta(x_2 - x_1)$ &
    $\delta(x_1 - x_2)$ \\ [0.3ex]
    \hline
    $\sA + \sB \rightleftharpoons \sC + \sD$ &
    \makecell[l]{
    $p \delta_{(x_1,x_2)}(x_3,x_4)$ \\
    $\qquad +(1-p) \delta_{(x_1,x_2)}(x_4,x_3)$}  &
    \makecell[l]{
    $p \delta_{(x_3,x_4)}(x_1,x_2)$ \\
    $\qquad +(1-p) \delta_{(x_3,x_4)}(x_2,x_1)$} \\ [1.5ex]
    \hline
    \end{tabular}}

	\vspace{3pt}
    \footnotesize
    \rev{Note, $\vx^{\vn^+}_{\va}$ and $\vx^{\vn^-}_{\vb}$ respectively denote
	the substrate positions for the forward and backward reactions.
	$\vx^{\vn}_{\vb}$ and $\vx^{\vn}_{\va}$ similarly denote the product
	positions (for more details see section \ref{s3-1}).}
\end{table}
\begin{table}[htbp]
	\centering
    \caption{Velocity Placement Densities (assuming conservation of mass)}
    \label{s1 tab: v placement}
    \makebox[\linewidth]{
    \begin{tabular}{| c | c | c |}
    \hline
    Reaction &
    $m_+^{\beta}(\vv^{\vn}_{\vb} | \vv^{\vnp}_{\va})$ &
    $m_-^{\beta}(\vv^{\vn}_{\va} | \vv^{\vnm}_{\vb})$ \\ [0.5ex]
    \hline
    $\sA + \sB \rightleftharpoons \sC$ &
    $\delta \Big(v_3 - \frac{m_1 v_1 + m_2 v_2}{m_3} \Big)$ &
    \makecell[l]{
    $\delta\paren{v_3 - \tfrac{(m_1 v_1 + m_2 v_2)}{m_3}}$ \\
    $\times \GaussianDensity_{d}(v_1-v_2;(D_1 \beta_1 + D_2 \beta_2)\mathrm{I}_{d})$
    } \\ [3ex]
    \hline
    $\sA \rightleftharpoons \sB$ &
    $\delta(v_2 - v_1)$ &
    $\delta(v_1 - v_2)$ \\ [0.3ex]
    \hline
    $\sA + \sB \rightleftharpoons \sC + \sD$ &
    \makecell[l]{
    $(m_3 + m_4)^d \delta\paren{\mathrm{p}_{3 4} - \mathrm{p}_{1 2}}$ \\
    $\times
    \GaussianDensity_{d}(v_3-v_4;(D_3\beta_3+D_4\beta_4)\mathrm{I}_{d})$
    } &
    \makecell[l]{
    $(m_1 + m_2)^d \delta\paren{\mathrm{p}_{1 2} - \mathrm{p}_{3 4}}$ \\
    $\times
    \GaussianDensity_{d}(v_1-v_2;(D_1 \beta_1 + D_2 \beta_2)\mathrm{I}_{d})$
    } \\ [3ex]
    \hline
    \end{tabular}}
	
	\vspace{3pt}
    \footnotesize
    \rev{Here $\mathrm{p_{ij}} = m_i v_i + m_j v_j$, $(i,j) \in \{(1,2),
	(3,4)\}$. Note, $\vv^{\vn^+}_{\va}$ and $\vv^{\vn^-}_{\vb}$ respectively
	denote the substrate velocities for the forward and backward reactions.
	$\vv^{\vn}_{\vb}$ and $\vv^{\vn}_{\va}$ similarly denote the product
	velocities (for more details see section \ref{s3-1}).}
\end{table}
\begin{remark}\label{s2 remark: scaling} In all three examples, in the
	non-dimensional coordinate system we find that $\beta$ factors out from the
	$\beta$-dependent velocity placement density $m^{\beta}_{\pm}$, only
	modulating its amplitude. The transformed densities $\widetilde{m}_{\pm}$
	are independent of $\beta$. Inspired by these observations, we now assume a
	generalization of this scaling property to study general reversible
	reactions. In \cref{s4}, we demonstrate that this scaling property plays a
	key role in deriving the overdamped limit of reactive Langevin Dynamics,
	enabling its consistency with overdamped models.
\end{remark}

\section{Formulation of Reactive Langevin Dynamics for General Reversible
Reactions}\label{s3} In this section, we first introduce additional notations
and definitions for modeling general reversible reactions. We then introduce the
detailed balance relation and assumptions about rate functions and placement
densities inspired from the examples in \cref{s2}, which are key components for
deriving the overdamped, $\beta \to \infty$, limit of RLD models in \cref{s4}.

\subsection{Preliminary Definitions}\label{s3-1}
Recall the generic reaction \cref{s2 eq: reaction}. Consider the population
vector $\vN(t) = \vn = (n_1, ..., n_J)$. We denote $\vnm$ as the population
state vector transitioned from the population state $\vn$ after a forward
reaction occurs, i.e.
	$\vn^- = \vn - \va + \vb$,
and, $\vnp$ as the population state vector transitioned from the population state $\vn$
after a backward reaction occurs
	$\vn^+ = \vn - \vb + \va$.

Next, we introduce a system of notations to encode substrate and particle states
and configurations that are needed to later specify reaction processes.
\begin{definition}\label{s3-1: def1}
	For the generic reversible reaction \cref{s2 eq: reaction}, let $\Ipp(\vn)$
	$\subset (\N \backslash \{0\})^{|\va|}$ denote the substrate index space of
	the forward reaction when $\vN(t) = \vn$. Denote by $\vec{i}^{\vn}_{\va} \in
	\Ipp(\vn)$ the indices for one possible set of substrates, i.e.
	\begin{equation*}
		\ina = \big(i^{(1)}_1, ..., i^{(1)}_{a_1}, ...,
					 i^{(J)}_1, ..., i^{(J)}_{a_J} \big).
	\end{equation*}
	Here, $i^{(j)}_l \in \{1, 2, ..., n_j\}$ labels the index of the $l$-th
	substrate particle of species $j$, and we assume $i^{(j)}_1 \leq i^{(j)}_2
	\leq \cdots i^{(j)}_{a_j}$ for all $j = 1, 2, ..., J$. The substrate index
	space $\Imm(\vn)$ and specific substrate indices, $\inb$, of the backward
	reaction can be defined analogously.
\end{definition}

\begin{definition}\label{s3-1: def2}
	For the generic reversible reaction \cref{s2 eq: reaction}, corresponding to the
	substrate index space $\Ipp(\vn)$, we can define the forward substrate state space
		$\Xi_+(\vxi^{\vn}) \subset (\Omega \times \R^d)^{|\va|}$.
	Denote $\vxi^{\vn}_{\va} \in \Xi_+(\vxi^{\vn})$ as the state vector for
	the set of substrate particles with indices $\ina$, so that
	\begin{equation*}
		\vxi^{\vn}_{\va}
		= \left( \vxi^{\vn}_{\va_1}, \vxi^{\vn}_{\va_2}, ..., \vxi^{\vn}_{\va_J} \right)
		=
		\Big(
			\Big( \vxi_{i^{(1)}_1}^{\vn}, ..., \vxi_{i^{(1)}_{a_1}}^{\vn} \Big),
			\Big( \vxi_{i^{(2)}_1}^{\vn}, ..., \vxi_{i^{(2)}_{a_2}}^{\vn} \Big)
			, ...,
			\Big( \vxi_{i^{(J)}_1}^{\vn}, ..., \vxi_{i^{(J)}_{a_J}}^{\vn} \Big)
		\Big).
	\end{equation*}
	\rev{Note that $\Xi_+(\vxi^{\vn})$ then represents the (finite) set of all
	valid substrate vectors that can be extracted from a specific state
	$\vxi^{\vn}$.} We can similarly define substrate position vectors,
	$\vx^{\vn}_{\va}$, and velocity vectors, $\vv^{\vn}_{\va}$. The backward
	reaction substrate state space $\Xi_-(\vxi^{\vn}) \subset (\Omega \times
	\R^d)^{|\vb|}$ and the sampled vectors $\xinb$, $\vx^{\vn}_{\vb}$, and
	$\vv^{\vn}_{\vb}$ are defined analogously.
\end{definition}

\subsection{Kolmogorov Forward Equation}
Recall $\vP(t) = \{p^{\vn}(\vxi^{\vn}, t)\}_{\vn}$ representing the collection of probability
densities over all possible states at time $t$. The evolution equation for each probability density
$p^{\vn}(\vxi^{\vn}, t)$, based on the dynamics \cref{s2 eq: Langevin Dynamics} and the generic
reaction \cref{s2 eq: reaction}, follows the Kolmogorov forward equation
\begin{equation}\label{s3 eq: Kolmogorov forward eq}
	\frac{\partial p^{\vn}}{\partial t} (\vxi^{\vn}, t)
	= (\Lp + \Rpp + \Rmm) p^{\vn}(\vxi^{\vn}, t),
\end{equation}
where the transport operator $\Lp$ is defined by
\begin{align}
	\Lp p^{\vn} (\vxi^{\vn}, t) \label{s3 eq: L}
	&= \Bigg( \sum_{j=1}^J \sum_{l=1}^{n_j} \Lp_{(x^{(j)}_l, v^{(j)}_l)} \Bigg)
	   p^{\vn} (\vxi^{\vn}, t) \\
	\Lp_{(x^{(j)}_l, v^{(j)}_l)} \label{s3 eq: Lp}
	&=  \beta_j \nabla_{v^{(j)}_l}
		\cdot [v^{(j)}_l + \beta_j D_j \nabla_{v^{(j)}_l}]
		- v^{(j)}_l \cdot \nabla_{x^{(j)}_l}.
\end{align}

To define the reaction operators, $\Rpp$ and $\Rmm$, we introduce notations for adding
or removing a particle from a given state $\vxi^{\vn}$.  Let
\begin{equation*}
	\vxi^{\vn} \cup \tilde{\xi}^{(j)}
	= (\vxi^{n_1}, ..., \vxi^{n_{j-1}},
		(\vxi^{n_j}, \tilde{\xi}^{(j)}),
		\vxi^{n_{j+1}}, ..., \vxi^{n_J})
\end{equation*}
represent adding a new particle of species $j$ with state $\tilde{\xi}^{(j)}$
into the current system $\vxi^{\vn}$. This notation can be naturally extended to
adding multiple particles in a system. For example, $\vxi^{\vn} \cup
\tilde{\vxi}$ denotes adding multiple particles with states given by the
combined vector $\tilde{\vxi}$ to a system $\vxi^{\vn}$. We use the notation
\begin{equation*}
	\vxi^{\vn} \backslash \xi^{n_j}_{l}
	= (\vxi^{n_1}, ...,
		(\xi^{n_j}_1, ..., \xi^{n_j}_{l-1}, \xi^{n_j}_{l+1}, ..., \xi^{n_j}_{n_j}),
		..., \vxi^{n_J})
\end{equation*}
to represent removing the $l$th particles of species $j$ in the system $\vxi^{\vn}$,
which can be also extended to removing multiple particle from a given system.

With these notations, the forward, $\sA + \sB \to \sC$, reaction operator,
$\Rpp$, is
\begin{equation}\label{s3 eq: Rpp}
\begin{multlined}
	\Rpp p^{\vn}(\vxi^{\vn}, t)
	=  - \Bigg( \sum_{\xina \in \Xi_+(\vxi^{\vn})}
				 K_{+}^{\beta} (\vxi^{\vn}_{\va})\Bigg)
			p^{\vn} (\vxi^{\vn}, t) \\
		+ \sum_{\xinb \in \Xi_-(\vxi^{\vn})}
		  \rev{\frac{1}{\va !}} \int_{(\Omega \times \R^d)^{|\va|}}
		m_{+}^{\beta} (\vxi^{\vn}_{\vb} | \vxi^{\vnp}_{\va})
		K_{+}^{\beta} (\vxi^{\vnp}_{\va})
		p^{\vnp}((\vxi^{\vn} \backslash \vxi^{\vn}_{\vb}) \cup \vxi^{\vnp}_{\va}, t)
		\, \mathrm{d} \vxi^{\vnp}_{\va},
\end{multlined}
\end{equation}
where the reaction rate function $K_{+}^{\beta} (\vxi^{\vn}_{\va})$ represents
the probability per time the substrates at $\vxi^{\vn}_{\va}$ react, \rev{the
integration variable $\vxi^{\vnp}_{\va}$ represents the positions of a possible
set of reaction substrates that upon reacting could produce products at
$\vxi^{\vn}_{\vb}$,} and the placement density $m_{+}^{\beta} (\vxi^{\vn}_{\vb}
| \vxi^{\vnp}_{\va})$ represents the probability density that products are
created at $\vxi^{\vn}_{\vb}$ from substrates at $\vxi^{\vnp}_{\va}$. The
superscript $\beta$ is to indicate functions which may depend on $\beta$.
\rev{Note that the $1 / \va!$ factor in the second line arises due to
overcounting of distinct substrate states within the integral, see \cref{rmk:
indistinguishability}.} We analogously define the backward, $\sC \to \sA + \sB$,
reaction operator $\Rmm$ as
\begin{equation}
\begin{multlined}
	\Rmm p^{\vn}(\vxi^{\vn}, t)
	=  - \Bigg(\sum_{\xinb \in \Xi_-(\vxi^{\vn})}
				K_{-}^{\beta} (\vxi^{\vn}_{\vb})\Bigg)
			p^{\vn} (\vxi^{\vn}, t) \\
	   + \sum_{\xina \in \Xi_+(\vxi^{\vn})}
	     \rev{\frac{1}{\vb !}} \int_{(\Omega \times \R^d)^{|\vb|}}
		m_{-}^{\beta} (\vxi^{\vn}_{\va} | \vxi^{\vnm}_{\vb})
		K_{-}^{\beta} (\vxi^{\vnm}_{\vb})
		p^{\vnm}((\vxi^{\vn} \backslash \vxi^{\vn}_{\va}) \cup \vxi^{\vnm}_{\vb}, t)
		\, \mathrm{d} \vxi^{\vnm}_{\vb}.\label{s3 eq: Rmm}
\end{multlined}
\end{equation}

\subsection{Abstract Detailed Balance Relation}
As in the over-damped case~\cite{ZhangIsaacson2022}, when the system is closed
(i.e. $\Omega$ is finite with a reflecting Neumann, or periodic, boundary
condition), at equilibrium the \textit{principle of detailed balance} should
hold for the pointwise reaction fluxes. That is, the equilibrium solutions
$\bar{\vP} = \{\bar{p}^{\vn}(\vxi^{\vn})\}_{\vn}$ should satisfy
\begin{align}
	m_{+}^{\beta} (\vxi^{\vn}_{\vb} | \vxi^{\vnp}_{\va})
	K_{+}^{\beta} (\vxi^{\vnp}_{\va})
	\bar{p}^{\vnp}((\vxi^{\vn} \backslash \vxi^{\vn}_{\vb}) \cup \vxi^{\vnp}_{\va})
	&= m_{-}^{\beta} (\vxi^{\vnp}_{\va} | \vxi^{\vn}_{\vb})
	K_{-}^{\beta} (\vxi^{\vn}_{\vb})
	\bar{p}^{\vn}(\vxi^{\vn}), \label{s3 eq: detailed balance 1}  \\
	m_{-}^{\beta} (\vxi^{\vn}_{\va} | \vxi^{\vnm}_{\vb})
	K_{-}^{\beta} (\vxi^{\vnm}_{\vb})
	\bar{p}^{\vnm}((\vxi^{\vn} \backslash \vxi^{\vn}_{\va}) \cup \vxi^{\vnm}_{\vb})
	&=
	m_{+}^{\beta} (\vxi^{\vnm}_{\vb} | \vxi^{\vn}_{\va})
	K_{+}^{\beta} (\vxi^{\vn}_{\va})
	\bar{p}^{\vn}(\vxi^{\vn}). \label{s3 eq: detailed balance 2}
\end{align}
Substituting \cref{s3 eq: detailed balance 1} into
the forward reaction operator \cref{s3 eq: Rpp}, we have
\begin{equation}\label{s3 eq: detailed balance in R plus}
\begin{aligned}
	\Rpp \bar{p}^{\vn}(\vxi^{\vn})
	=
	\Bigg(
		- \sum_{\xina \in \Xi_+(\vxi^{\vn})} K_{+}^{\beta} (\vxi^{\vn}_{\va})
		+ \sum_{\xinb \in \Xi_-(\vxi^{\vn})} K_{-}^{\beta} (\vxi^{\vn}_{\vb})
	\Bigg)
	\bar{p}^{\vn}(\vxi^{\vn}) .
\end{aligned}
\end{equation}
Similarly, substituting (\ref{s3 eq: detailed balance 2}) into (\ref{s3 eq: Rmm}) gives
\begin{equation}\label{s3 eq: detailed balance in R minus}
\begin{aligned}
	\Rmm \bar{p}^{\vn}(\vxi^{\vn})
	=
	\Bigg(
		- \sum_{\xinb \in \Xi_-(\vxi^{\vn})} K_{-}^{\beta} (\vxi^{\vn}_{\vb})
		+ \sum_{\xina \in \Xi_+(\vxi^{\vn})} K_{+}^{\beta} (\vxi^{\vn}_{\va})
	\Bigg)
	\bar{p}^{\vn}(\vxi^{\vn}).
\end{aligned}
\end{equation}
Combining (\ref{s3 eq: detailed balance in R plus}) and (\ref{s3 eq: detailed balance in R minus}),
we have $(\Rpp + \Rmm) \bar{p}^{\vn}(\vxi^{\vn}) = 0$, which implies that
\begin{equation}\label{s3 eq: Lp=0}
	\Lp \bar{p}^{\vn}(\vxi^{\vn}) = 0.
\end{equation}
The appropriate equilibrium solution of equation (\ref{s3 eq: Lp=0}), coming
from the long-time behavior in the absence of reactions, is a uniform
distribution in space and Maxwell-Boltzmann distribution in velocity
\begin{equation}\label{s3 eq: limit density}
	\bar{p}^{\vn} (\vxi^{\vn})
	= \frac{\vn ! \pi(\vn)}{|\Omega|^{|\vn|}}
	  \prod_{j=1}^J \prod_{l=1}^{n_j}
      \GaussianDensity_{d}(v^{(j)}_l;(D_j \beta_j)\mathrm{I}_{d}),
\end{equation}
where $\pi(\vn)$ denotes the equilibrium probability to have the population state $\vn$, i.e.
\begin{equation*}
	\pi(\vn) = \lim_{t \to \infty} \mathbb{P}(\vN(t) = \vn) = \frac{1}{\vn !} \int_{(\Omega \times \R^{d})^{|\vn|}}
	\bar{p}^{\vn} (\vxi^{\vn}) \mathrm{d} \vxi^{\vn}.
\end{equation*}

Let $\Kd$ denote the equilibrium dissociation constant of the reaction. As the
system is spatially well-mixed at equilibrium, $\pi(\vn)$ satisfies the
corresponding non-spatial, well-mixed, equilibrium chemical master equation
model, from which we have that $\Kd = |\Omega|^{(|\vb| - |\va|)} \frac{\vnp\!!\,
\pi(\vnp)}{\vn! \, \pi(\vn)}$ (see \cite{van1976equilibrium}).
Substituting the equilibrium solution (\ref{s3 eq: limit density}) into the
detailed balance relation (\ref{s3 eq: detailed balance 1}) then gives
\begin{equation} \label{s3 eq: detailed balance 3}
	\begin{multlined}
		\Kd m_{+}^{\beta} (\vxi^{\vn}_{\vb} | \vxi^{\vnp}_{\va})
		 K_{+}^{\beta} (\vxi^{\vnp}_{\va})\prod_{j=1}^J
            \GaussianDensity_{a_{j}d}(\vv^{\vnp}_{\va_j};(D_j \beta_j)\mathrm{I}_{a_j d}) \\
		= m_{-}^{\beta} (\vxi^{\vnp}_{\va} | \vxi^{\vn}_{\vb})
		K_{-}^{\beta} (\vxi^{\vn}_{\vb}) \prod_{j=1}^J
		\GaussianDensity_{b_{j}d}(\vv^{\vn}_{\vb_j};(D_j \beta_j)\mathrm{I}_{b_j d}).
	\end{multlined}
\end{equation}

\subsection{Assumptions on Reaction Functions and Placement Densities}\label{s3: assumptions}

Motivated by the examples in \cref{s2}, we make the following assumptions
regarding the rate functions and placement densities.
\begin{assumption}\label{s3 assump1}
	The reaction rate functions only depend on
	positions and are independent of the friction constant $\beta$: $K_{+}^{\beta} (\vxi^{\vn}_{\va}) = K_{+} (\vx^{\vn}_{\va})$ and $
		K_{-}^{\beta} (\vxi^{\vn}_{\vb}) = K_{-} (\vx^{\vn}_{\va})$.
\end{assumption}

\begin{assumption}\label{s3 assump2} Each placement density can be decomposed
	into a product of two placement densities, one depending on positions and
	the other on velocities only, i.e.
	\begin{align*}
		m_{+}^{\beta} (\vxi^{\vn}_{\vb} | \vxi^{\vnp}_{\va})
		&=
		m_{+} (\vx^{\vn}_{\vb} | \vx^{\vnp}_{\va})
		m_{+}^{\beta} (\vv^{\vn}_{\vb} | \vv^{\vnp}_{\va}), \\
		m_{-}^{\beta} (\vxi^{\vnp}_{\va} | \vxi^{\vn}_{\vb}),
		&=
		m_{-} (\vx^{\vnp}_{\va} | \vx^{\vn}_{\vb})
		m_{-}^{\beta} (\vv^{\vnp}_{\va} | \vv^{\vn}_{\vb}).
	\end{align*}
	As we mentioned in \cref{s2 remark: m}, here we use the same notation
	$m_{\pm}^{\beta}(\cdot|\cdot)$ to represent the probability density of the
	first argument given the second one, regardless of whether these arguments
	pertain to particle position, velocity, or state. Additionally, we assume
	that the placement densities of positions are independent of $\beta$,
	\rev{and hence drop the $\beta$ superscript for them}.

\end{assumption}

\begin{assumption}\label{s3 assump3}
	We assume all placement densities are probability densities, which are
	non-negative and can be integrated to one, i.e.
	\begin{align*}
		  \rev{\frac{1}{\vb !}} \int_{\Omega^{|\vb|}} m_{+} (\vx^{\vn}_{\vb} | \vx^{\vnp}_{\va}) \mathrm{d} \vx^{\vn}_{\vb}
		&= \rev{\frac{1}{\va !}} \int_{\Omega^{|\va|}} m_{-} (\vx^{\vnp}_{\va} | \vx^{\vn}_{\vb}) \mathrm{d} \vx^{\vnp}_{\va} = 1 ,\\
			\int_{\R^{|\vb| d}} m_{+}^{\beta} (\vv^{\vn}_{\vb} | \vv^{\vnp}_{\va}) \mathrm{d} \vv^{\vn}_{\vb}
		&=  \int_{\R^{|\va| d}} m_{-}^{\beta} (\vv^{\vnp}_{\va} | \vv^{\vn}_{\vb}) \mathrm{d} \vv^{\vnp}_{\va} = 1.
	\end{align*}
\end{assumption}
\begin{remark} \label{rmk: indistinguishability} \rev{Note, the scaling terms
	$\frac{1}{\va !}$ and $\frac{1}{\vb !}$ multiplying the integrals with
	respect to position are due to the indistinguishability of particles of the
	same species. They ensure that each unique state is counted only once in
	reactions that involve multiple substrates or products of the same species,
	such as $2\sA \to \sB$. While one could alternatively modify the state space
	to account for this indistinguishability, e.g. by assuming an ordering of
	the positions (or the velocities) of particles of the same species, see for
	example \cite{DoiSecondQuantA,DoiSecondQuantB}, we find it more convenient
	to instead include such rescalings. Note, it is only necessary to rescale or
	order one of position or velocity to avoid overcounting of combined
	position-velocity states. We choose to rescale the position integrals since
	we will subsequently project out the velocity components and focus on the
	position dynamics when taking the overdamped limit in \cref{s4}.}
\end{remark}
\begin{assumption}\label{s3 assump4} Consider the non-dimensional coordinates
	$v^{(j)}_{l} = \sqrt{\beta\gamma_j} \eta^{(j)}_{l}$. Based on the
	observations in \cref{s2}, we assume that the velocity placement densities
	have the following scalings in $\beta$ when non-dimensionalized
	\begin{align*}
		m_{+}^{\beta} (\vv^{\vn}_{\vb} | \vv^{\vnp}_{\va})
 		&=
		\Bigg(\prod_{j=1}^J \frac{1}{(\beta \gamma_j)^{b_j d/2}}\Bigg)
		\widetilde{m}_+ (\veta^{\vn}_{\vb} \mid \veta^{\vnp}_{\va}), \\
		m_{-}^{\beta} (\vv^{\vn}_{\va} | \vv^{\vnm}_{\vb})
		&=
		\Bigg(\prod_{j=1}^J \frac{1}{(\beta \gamma_j)^{a_j d/2}}\Bigg)
		\widetilde{m}_- (\veta^{\vn}_{\va} \mid \veta^{\vnm}_{\vb}).
	\end{align*}
	Note, \cref{s3 assump3} then implies that $\widetilde{m}_+ (\veta^{\vn}_{\vb}
	\mid \veta^{\vnp}_{\va})$ and $\widetilde{m}_- (\veta^{\vn}_{\va} \mid
	\veta^{\vnm}_{\vb})$ are normalized densities in $\veta^{\vn}_{\vb}$ and
	$\veta^{\vn}_{\va}$ respectively.
\end{assumption}

Let us now revisit the detailed balance relation (\ref{s3 eq: detailed balance
1}). Considering the assumptions above, we integrate both sides of equation
(\ref{s3 eq: detailed balance 3}) against $\vv^{\vn}_{\vb}$ and
$\vv^{\vnp}_{\va}$ and get
\begin{equation}\label{s3 eq: overdamped detailed balance}
	\Kd m_{+}(\vx^{\vn}_{\vb} | \vx^{\vnp}_{\va}) K_{+} (\vx^{\vnp}_{\va})
	= m_{-} (\vx^{\vnp}_{\va} | \vx^{\vn}_{\vb}) K_{-} (\vx^{\vn}_{\vb}),
\end{equation}
which is the detailed balance relation of the overdamped model
\cite{ZhangIsaacson2022}. Using (\ref{s3 eq: overdamped detailed balance}) to simplify (\ref{s3 eq: detailed balance 3}), and converting to non-dimensional velocity coordinates, we get
\begin{equation}\label{s3 eq: detailed balance 4 eta}
\begin{aligned}
	(2\pi)^{\sum_{j=1}^J(b_j - a_j)d/2}
	\exp \left(\tfrac{|\veta^{\vn}_{\vb}|^2 - |\veta^{\vnp}_{\va}|^2}{2} \right)
	\widetilde{m}_{+} (\veta^{\vn}_{\vb} | \veta^{\vnp}_{\va})
	&=	\widetilde{m}_{-} (\veta^{\vnp}_{\va} | \veta^{\vn}_{\vb}).
\end{aligned}
\end{equation}
Integrating with respect to $\veta^{\vnp}_{\va}$, we find the identity that
\begin{equation}\label{s3 eq: integration property 1}
	(2\pi)^{\sum_{j=1}^J(b_j - a_j)d/2}
	\int_{\R^{|\va|d}}
	\exp \left(\tfrac{|\veta^{\vn}_{\vb}|^2 - |\veta^{\vnp}_{\va}|^2}{2} \right)
	\widetilde{m}_{+} (\veta^{\vn}_{\vb} | \veta^{\vnp}_{\va})
	\, \mathrm{d} \rev{\veta^{\vn^+}_{\va}}
	= 1.
\end{equation}
Similarly, using the detailed balance relation (\ref{s3 eq: detailed balance
2}),  the same procedure gives
\begin{equation}\label{s3 eq: integration property 2}
	(2\pi)^{\sum_{j=1}^J(a_j - b_j)d/2}
	\int_{\R^{|\vb|d}}
	\exp \left(\tfrac{|\veta^{\vn}_{\va}|^2 - |\veta^{\vnm}_{\vb}|^2}{2} \right)
	\widetilde{m}_{-} (\veta^{\vn}_{\va} | \veta^{\vnm}_{\vb})
	\, \mathrm{d} \veta^{\vn^-}_{\vb}
	= 1.
\end{equation}
It can be verified that \eqref{s3 eq: integration property 1}-\eqref{s3 eq:
integration property 2} hold for each example in \cref{s2}.
As we will see in the next section, these identities are key components in
ensuring reaction terms have the right order in $\beta$ so that we recover the
over-damped reaction model as $\beta \to \infty$.

\section{Overdamped Limit of Reactive Langevin Dynamics}\label{s4}
In this section, we show via \rev{formal} asymptotic expansion that the overdamped limit, $\beta \to
\infty$, of the solution to the RLD model \cref{s3 eq: Kolmogorov forward eq} is
the solution to the VR PBSRD model.

\subsection{\cref{s3 eq: Kolmogorov forward eq} in Non-Dimensionalized Variables}\label{s4: transformations}
Let
\begin{equation*}
	\bar{u}^{\vn}(\vv^{\vn})
	= \prod_{j=1}^J \prod_{l=1}^{n_j}\GaussianDensity_{d}(v^{(j)}_l;(D_j\beta_j)\mathrm{I}_{d})
\end{equation*}
%
denote the Maxwell-Boltzmann velocity distributions associated with the velocity equilibria components
in the non-reactive case. We factor
\begin{equation}\label{s4 eq: factorize p}
	p^{\vn}(\vxi^{\vn}, t)
		:= \bar{u}^{\vn}(\vv^{\vn}) \rho^{\vn}(\vxi^{\vn}, t).
\end{equation}

To substitute the factorization (\ref{s4 eq: factorize p}) into the forward Kolmogorov
equation (\ref{s3 eq: Kolmogorov forward eq}), let us first consider the transport operator $\Lp$.
For each summand of $\Lp$, we have
\begin{align*}
	  \Lp_{(x^{(j)}_l, v^{(j)}_l)} p^{\vn}(\vxi^{\vn}, t)
	&= \bar{u}^{\vn}(\vv^{\vn})
		\left[
			\left(
		  	\Lp^{(1)}_{v^{(j)}_l} +
			\Lp^{(2)}_{(x^{(j)}_l, v^{(j)}_l)}
			\right)
			\rho^{\vn}(\vxi^{\vn}, t)
		\right],
\end{align*}
where $\Lp^{(1)}_{v^{(j)}_l}
	= \beta_j (\beta_j D_j \Delta_{v^{(j)}_l} -v^{(j)}_l \cdot \nabla_{v^{(j)}_l})$ and
	$\Lp^{(2)}_{(x^{(j)}_l, v^{(j)}_l)}
	= - v^{(j)}_l \cdot \nabla_{x^{(j)}_l}$.
Hence, the transport operator becomes
\begin{align*}
	   \Lp p^{\vn}(\vxi^{\vn}, t)
	 & =\bar{u}^{\vn} (\vv^{\vn})
		\left(\Lp^{(1)} + \Lp^{(2)}\right)
		\rho^{\vn}(\vxi^{\vn}, t) ,
\end{align*}
where we denote $\Lp^{(1)} = \sum_{j=1}^J \sum_{l=1}^{n_j} \Lp^{(1)}_{v^{(j)}_l}$ and $\Lp^{(2)} = \sum_{j=1}^J \sum_{l=1}^{n_j} \Lp^{(2)}_{(x^{(j)}_l, v^{(j)}_l)}$.
With these definitions, the Kolmogorov forward equation (\ref{s3 eq: Kolmogorov forward eq}) transforms
to
\begin{equation}\label{s4 eq: Kolmogorov rho}
\begin{aligned}
		\frac{\partial \rho^{\vn}}{\partial t} (\vxi^{\vn}, t)
	  &= \big(\Lp^{(1)} + \Lp^{(2)} + \Rpp^\dagger + \Rmm^\dagger \big) \rho^{\vn}(\vxi^{\vn}, t) \\
	  &\quad - \Bigg(
					  \sum_{\vx^{\vn}_{\va} \in \Xi_+(\vx^{\vn})} K_{+} (\vx^{\vn}_{\va})
					+ \sum_{\vx^{\vn}_{\vb} \in \Xi_-(\vx^{\vn})} K_{-} (\vx^{\vn}_{\vb})
				\Bigg)
	  		\rho^{\vn}(\vxi^{\vn}, t) ,
\end{aligned}
\end{equation}
\rev{where, by Assumption~\ref{s3 assump1}, we assume that the functions
$K_+(\cdot)$ and $K_-(\cdot)$ depend only on the positions of the particles.
Similar to Definition~\ref{s3-1: def2}, we define the forward substrate position
space $\Xi_{+}(\vx^{\vn}) \subset \Omega^{|\va|}$, and denote by
$\vx^{\vn}_{\va} \in \Xi_{+}(\vx^{\vn})$ the position vector corresponding to
the set of substrate particles with indices in $\ina$. Analogously, we define
the backward substrate position space $\Xi_{-}(\vx^{\vn}) \subset
\Omega^{|\vb|}$ and the corresponding subvector $\vx^{\vn}_{\vb}$. The operators
$\Rpp^\dagger$ and $\Rmm^\dagger$ acting on the density $\rho^{\vn}(\vxi^{\vn},
t)$ then take the following form}
\begin{equation*}
\begin{aligned}
	\Rpp^\dagger \rho^{\vn}(\vxi^{\vn}, t)
	&:= \frac{1}{\bar{u}^{\vn} (\vv^{\vn})}
		\sum_{\xinb \in \Xim(\vxi^{\vn})}
		\rev{\frac{1}{\va !}} \int_{(\Omega \times \R^d)^{|\va|}}
			m_{+}^{\beta} (\vxi^{\vn}_{\vb} | \vxi^{\vnp}_{\va})
			K_{+} (\vx^{\vnp}_{\va}) \\
	& \qquad \qquad
		\bar{u}^{\vnp}((\vv^{\vn} \backslash \vv^{\vn}_{\vb}) \cup \vv^{\vnp}_{\va})
		\rho^{\vnp}((\vxi^{\vn} \backslash \vxi^{\vn}_{\vb}) \cup \vxi^{\vnp}_{\va}, t)
		\mathrm{d} \vxi^{\vnp}_{\va},
\end{aligned}
\end{equation*}
and
\begin{equation*}
\begin{aligned}
	\Rmm^\dagger \rho^{\vn}(\vxi^{\vn}, t)
	&:= \frac{1}{\bar{u}^{\vn} (\vv^{\vn})}
		\sum_{\xina \in \Xip(\vxi^{\vn})}
		\rev{\frac{1}{\vb !}} \int_{(\Omega \times \R^d)^{|\vb|}}
		m_{-}^{\beta} (\vxi^{\vn}_{\va} | \vxi^{\vnm}_{\vb})
		K_{-} (\vx^{\vnm}_{\vb}) \\
	& \qquad \qquad
		\bar{u}^{\vnm}((\vv^{\vn} \backslash \vv^{\vn}_{\va}) \cup \vv^{\vnm}_{\vb})
		\rho^{\vnm}((\vxi^{\vn} \backslash \vxi^{\vn}_{\va}) \cup \vxi^{\vnm}_{\vb}, t)
		\mathrm{d} \vxi^{\vnm}_{\vb}.
\end{aligned}
\end{equation*}

Assume that $\beta_j = \beta \hat{\beta}_{j}$ and define $\gamma_j = D_j \hat{\beta}_j$.
Analogously to \cite{IsaacsonErban2016}, we introduce non-dimensional velocities,
$v^{(j)}_l = \sqrt{\beta_j D_j} \eta_l^{(j)} = \sqrt{\beta \gamma_j} \eta^{(j)}_l$.
In the new coordinates, we denote the rescaled transport operators by
\begin{align}
	 \Lp^{(1)}_{v^{(j)}_l}  &\rightarrow \beta \hat{\Lp}^{(1)}_{\eta^{(j)}_l}, &
	  \hat{\Lp}^{(1)}_{\eta^{(j)}_l} &:= \hat{\beta}_j (\Delta_{\eta^{(j)}_l} - \eta^{(j)}_l \cdot \nabla_{\eta^{(j)}_l}) , \\
	\Lp^{(2)}_{(x^{(j)}_l, v^{(j)}_l)} &\rightarrow \sqrt{\beta} \hat{\Lp}^{(2)}_{(x^{(j)}_l, \eta^{(j)}_l)}, \label{s4 eq: hat L 2}
	  & \hat{\Lp}^{(2)}_{(x^{(j)}_l, \eta^{(j)}_l)} &:= -(\sqrt{\gamma_j})\ \eta^{(j)}_l \cdot \nabla_{x^{(j)}_l} ,
\end{align}
and define $\hat{\Lp}^{(1)} =
		\sum_{j=1}^J \sum_{l=1}^{n_j}
		\hat{\Lp}^{(1)}_{\eta^{(j)}_l}$ and $\hat{\Lp}^{(2)} =
		\sum_{j=1}^J \sum_{l=1}^{n_j}
		\hat{\Lp}^{(2)}_{(x^{(j)}_l, \eta^{(j)}_l)}$.

Let $\vzeta^{\vn} := (\vx^{\vn}, \veta^{\vn})$ and $f^{\vn}(\vzeta^{\vn}, t) := \rho^{\vn}(\vxi^{\vn}, t)$.
Using Assumption \ref{s3 assump4}
we have
\begin{align}
	\Rpp^\dagger \rho^{\vn}(\vxi^{\vn}, t)
	&=	\frac{1}{\bar{u}^{\vn} (\vv^{\vn})}
		\sum_{\xinb \in \Xim(\vxi^{\vn})}
		\rev{\frac{1}{\va !}} \int_{(\Omega \times \R^d)^{|\va|}}
		m_{+} (\vx^{\vn}_{\vb} | \vx^{\vnp}_{\va})
		K_{+} (\vx^{\vnp}_{\va}) \notag \\
	&\phantom{=} \quad \times
		\bar{u}^{\vnp}((\vv^{\vn} \backslash \vv^{\vn}_{\vb}) \cup \vv^{\vnp}_{\va})
		\rho^{\vnp}((\vxi^{\vn} \backslash \vxi^{\vn}_{\vb}) \cup \vxi^{\vnp}_{\va}, t)
		\ m_{+}^{\beta} (\vv^{\vn}_{\vb} | \vv^{\vnp}_{\va})
		\mathrm{d} \vxi^{\vnp}_{\va} \notag \\
	&=  (2\pi)^{\sum_{j=1}^J (b_j - a_j) d/2}
		\sum_{\zetanb \in \Xim(\vzeta^{\vn})}
		\rev{\frac{1}{\va !}} \int_{(\Omega \times \R^d)^{|\va|}}
			m_{+} (\vx^{\vn}_{\vb} | \vx^{\vnp}_{\va})
			K_+ (\vx^{\vnp}_{\va})  \notag \\
  	&\phantom{=} \quad \times
		\exp \left( \tfrac{|\veta^{\vn}_{\vb}|^2 - |\veta^{\vnp}_{\va}|^2}{2}\right)
  		f^{\vnp}((\vzeta^{\vn} \backslash \vzeta^{\vn}_{\vb}) \cup \vzeta^{\vnp}_{\va}, t)
		\ \widetilde{m}_{+} (\veta^{\vn}_{\vb} | \veta^{\vnp}_{\va})
		\mathrm{d} \vzeta^{\vnp}_{\va} \notag \\
	=:& \Rpp^\dagger[f^{\vnp}] (\vzeta^{\vn}, t) \label{s4 eq: R plus f}
\end{align}
and similarly, we have
\begin{align}
	\Rmm^\dagger \rho^{\vn}(\vxi^{\vn}, t)
	=&  (2\pi)^{\sum_{j=1}^J (a_j - b_j) d/2}
		\sum_{\zetana \in \Xip(\vzeta^{\vn})}
		\rev{\frac{1}{\vb !}} \int_{(\Omega \times \R^d)^{|\vb|}}
			m_{-} (\vx^{\vn}_{\va} | \vx^{\vnm}_{\vb})
			K_{-} (\vx^{\vnm}_{\vb}) \nonumber\\
  	& \times
		\exp \left( \tfrac{|\veta^{\vn}_{\va}|^2 - |\veta^{\vnm}_{\vb}|^2}{2}\right)
  		f^{\vnm}((\vzeta^{\vn} \backslash \vzeta^{\vn}_{\va}) \cup \vzeta^{\vnm}_{\vb}, t)
		\ \widetilde{m}_{-} (\veta^{\vn}_{\va} | \veta^{\vnm}_{\vb})
		\mathrm{d} \vzeta^{\vnm}_{\vb} \nonumber\\
	=:& \Rmm^\dagger[f^{\vnm}] (\vzeta^{\vn}, t) , \label{s4 eq: R minus f}
\end{align}
\rev{where, as before, $\Xip(\vzeta^{\vn})$ and $\Xim(\vzeta^{\vn})$ denote
the forward and backward substrate state spaces corresponding to the transformed state $\vzeta^{\vn}$, respectively.}
Note the key property that both reaction operators are now $O(1)$ in $\beta$.

Using the transformed operators, the forward equation (\ref{s4 eq: Kolmogorov rho})
becomes
\begin{equation}\label{s4 eq: Kolmogorov f}
\begin{aligned}
			\frac{\partial f^{\vn}}{\partial t} (\vzeta^{\vn}, t)
		  &= \big( \beta \hat{\Lp}^{(1)} + \sqrt{\beta} \hat{\Lp}^{(2)} \big)
		  	 f^{\vn}(\vzeta^{\vn}, t)
			 + \Rpp^\dagger[f^{\vnp}] (\vzeta^{\vn}, t)
			 + \Rmm^\dagger[f^{\vnm}] (\vzeta^{\vn}, t) \\
		  &\quad
		  	- \Bigg(
				\sum_{\vx^{\vn}_{\va} \in \Xip(\vx^{\vn})} K_{+} (\vx^{\vn}_{\va})
				+ \sum_{\vx^{\vn}_{\vb} \in \Xim(\vx^{\vn})} K_{-} (\vx^{\vn}_{\vb})
			  \Bigg)
				f^{\vn}(\vzeta^{\vn}, t),
\end{aligned}
\end{equation}

\subsection{Overdamped, $\beta \to \infty$, limit}\label{s4: overdamped limit}
We now develop an asymptotic expansion of $f^{\vn}$
as $\beta \rightarrow \infty$ of the form
\begin{equation}\label{s4 eq: expansion}
	f^{\vn}(\vzeta^{\vn}, t)
	\sim
	f^{\vn}_0 (\vzeta^{\vn}, t)
	+ \frac{1}{\sqrt{\beta}} f^{\vn}_1 (\vzeta^{\vn}, t)
	+ \frac{1}{\beta} f^{\vn}_2 (\vzeta^{\vn}, t)
	+ \cdots,
\end{equation}
Substituting the expansion into the forward equation (\ref{s4 eq: Kolmogorov f})
and equating terms of the same order in $\beta$, we find
\begin{align}
	\label{s4 eq: O beta}
		O(\beta): \qquad
		\hat{\Lp}^{(1)} f^{\vn}_0 &= 0, \\
	\label{s4 eq: O sqrt beta}
		O(\sqrt{\beta}): \quad
		- \hat{\Lp}^{(1)} f^{\vn}_1
		&= \hat{\Lp}^{(2)} f^{\vn}_0, \\
	\label{s4 eq: O 1}
		O(1): \quad
		- \hat{\Lp}^{(1)} f^{\vn}_2
		&= 	  \hat{\Lp}^{(2)} f^{\vn}_1
			- \frac{\partial f^{\vn}_0}{\partial t}
			+ \Rpp^\dagger[f^{\vnp}_0] (\vzeta^{\vn}, t) + \Rmm^\dagger[f^{\vnm}_0] (\vzeta^{\vn}, t) \\ \nonumber
		& \qquad - \Bigg(
					\sum_{\vx^{\vn}_{\va} \in \Xi_+(\vx^{\vn})} K_{+} (\vx^{\vn}_{\va})
		  			+ \sum_{\vx^{\vn}_{\vb} \in \Xi_-(\vx^{\vn})} K_{-} (\vx^{\vn}_{\vb})
					\Bigg)
			f^{\vn}_0 .
\end{align}

At $O(\beta)$, since the operator $\hat{\Lp}^{(1)}$ only depends on $\veta$, and
 represents the generator of a standard Ornstein-Uhlenbeck (OU) process,
 analogous to the expansions in~\cite{PavliotisStuartPhysD} we have that
 $f^{\vn}_0(\cdot)$ is a function depending only on position $\vx$, i.e.
\begin{equation*}
	f^{\vn}_0 (\vzeta^{\vn}, t)
	= g^{\vn} (\vx^{\vn}, t),
\end{equation*}
for some $g^{\vn}(\cdot)$. Likewise, $\hat{\Lp}^{(1)}$ has an associated
invariant density $\rho^\infty$ satisfying
\begin{equation}\label{s4 eq: L star rho = 0}
	\hat{\Lp}^{(1)^*} \rho^\infty = 0.
\end{equation}
Here, $\hat{\Lp}^{(1)^*}$ is the adjoint operator of $\hat{\Lp}^{(1)}$ with the
following form
\begin{align*}
	\hat{\Lp}^{(1)^*}
	&= \sum_{j=1}^J \sum_{l=1}^{n_j}
	\hat{\Lp}^{(1)^*}_{\eta^{(j)}_l},
	&\text{where  } &&
	\hat{\Lp}^{(1)^*}_{\eta^{(j)}_l} &= \hat{\beta}_j \nabla_{\eta^{(j)}_l} \cdot [\veta^{(j)}_l + \nabla_{\eta^{(j)}_l}].
\end{align*}
The velocity normalized invariant density solving (\ref{s4 eq: L star rho = 0}) is then
the Maxwell-Boltzmann distribution $
\rho^\infty (\veta^{\vn}) = \GaussianDensity(\veta^{\vn}; \mathrm{I}_{d}) \propto \exp \paren{-\tfrac{1}{2} \abs{\veta^{\vn}}^2}$.

Continuing with the expansion in $\beta$, the $O(\sqrt{\beta})$ equation (\ref{s4 eq: O
sqrt beta}) becomes
\begin{equation}\label{s4 eq: O beta 2}
	- \hat{\Lp}^{(1)} f^{\vn}_1 = \hat{\Lp}^{(2)} g^{\vn}.
\end{equation}
By the solvability condition for Poisson equations,
see~\cite{PardouxVeretennikov2001,PavliotisStuartPhysD}, (\ref{s4 eq: O beta 2})
has a solution if
\begin{equation}\label{s4 eq: sol cond}
	\int_{\R^{|\vn| d}}
	\hat{\Lp}^{(2)} g^{\vn} (\vx^{\vn}, t)
	\cdot \rho^\infty (\veta^{\vn})
	\mathrm{d} \veta^{\vn}
	= 0.
\end{equation}
We find the solvability condition (\ref{s4 eq: sol cond}) holds as the
velocity components of the integrand, $\eta_l^{(j)} \rho^\infty(\veta^{\vn})$,
are odd functions of $\eta_l^{(j)}$. Furthermore, we can find an
explicit solution of (\ref{s4 eq: O beta 2}) as
\begin{equation*}
	f_1^{\vn} (\vzeta^{\vn}, t)
	= - \sum_{j=1}^J \frac{\sqrt{\gamma_j}}{\hat{\beta}_j}
		\sum_{l=1}^{n_j}
		\left(
			\veta^{(j)}_l \cdot \nabla_{x^{(j)}_l}
			g^{\vn}(\vx^{\vn}, t)
		\right)
	   + \chi^{\vn} (\vx^{\vn}, t)
\end{equation*}
for some function $\chi^{\vn} (\vx^{\vn}, t)$.

For the $O(1)$ equation, \cref{s4 eq: O 1}, to be well posed, we again need the
solvability condition that the right side of \cref{s4 eq: O 1} is orthogonal to
the invariant measure, i.e.
\begin{equation}\label{s4 eq: sol cond O(1)}
\begin{aligned}
	0 &= \int_{\R^{|\vn| d}} \bigg[
	  \hat{\Lp}^{(2)} f^{\vn}_1
			- \frac{\partial g^{\vn}}{\partial t}
			+ \Rpp^\dagger[g^{\vnp}] (\vx^{\vn}, t)
			+ \Rmm^\dagger[g^{\vnm}] (\vx^{\vn}, t) \\
	&\phantom{=} \qquad \qquad
		- \bigg(
			\sum_{\vx^{\vn}_{\va} \in \Xi_+(\vx^{\vn})} K_{+} (\vx^{\vn}_{\va})
		  + \sum_{\vx^{\vn}_{\vb} \in \Xi_-(\vx^{\vn})} K_{-} (\vx^{\vn}_{\vb})
		 \bigg) g^{\vn} \bigg]
	  \rho^{\infty}(\veta^{\vn}) \, \mathrm{d} \veta^{\vn},
\end{aligned}
\end{equation}
where, by the integration properties (\ref{s3 eq: integration property 1}) and
(\ref{s3 eq: integration property 2}),
\begin{equation}\label{s4 eq: R plus g}
\begin{aligned}
		\Rpp^\dagger[g^{\vnp}] (\vx^{\vn}, t)
	& = \hspace{-1.2em} \sum_{\vx^{\vn}_{\vb} \in \Xim(\vx^{\vn})}
		\hspace{-.3em} \rev{\frac{1}{\va !}} \int_{\Omega^{|\va|}}
		m_{+} (\vx^{\vn}_{\vb} | \vx^{\vnp}_{\va})
		K_+ (\vx^{\vnp}_{\va})
		g^{\vnp} \hspace{-0.2em}((\vx^{\vn} \backslash \vx^{\vn}_{\vb}) \cup \vx^{\vnp}_{\va}\hspace{-0.3em}, t)
		\mathrm{d} \vx^{\vnp}_{\va}, \\
		\Rmm^\dagger[g^{\vnm}] (\vx^{\vn}, t)
	& = \hspace{-1.2em} \sum_{\vx^{\vn}_{\va} \in \Xip(\vx^{\vn})}
		\hspace{-.3em} \rev{\frac{1}{\vb !}} \int_{\Omega^{|\vb|}}
		m_{-} (\vx^{\vn}_{\va} | \vx^{\vnm}_{\vb})
		K_{-} (\vx^{\vnm}_{\vb})
		g^{\vnm} \hspace{-0.2em}((\vx^{\vn} \backslash \vx^{\vn}_{\va}) \cup \vx^{\vnm}_{\vb}\hspace{-0.3em}, t)
		\mathrm{d} \vx^{\vnm}_{\vb}.
\end{aligned}
\end{equation}
To simplify the solvability condition (\ref{s4 eq: sol cond O(1)}), we first
simplify $\hat{\Lp}^{(2)} f^{\vn}_1$. We have that
\begin{align*}
	&\hat{\Lp}^{(2)}_{(x^{(i)}_k, \eta^{(i)}_k)} f^{\vn}_1
	=
	-  	\sum_{j=1}^J
		\frac{\sqrt{\gamma_j}}{\hat{\beta}_j}
		\sum_{l=1}^{n_{j}}
		\hat{\Lp}^{(2)}_{(x^{(i)}_k, \eta^{(i)}_k)}
		\left(
			\veta^{(j)}_l \cdot \nabla_{x^{(j)}_l}
			g^{\vn}(\vx^{\vn}, t)
		\right)
	+  \hat{\Lp}^{(2)}_{(x^{(i)}_k, \eta^{(i)}_k)}
	   \chi^{\vn} (\vx^{\vn}, t) \\
	&\quad=
		\sum_{j=1}^J
		\frac{\sqrt{\gamma_j}}{\hat{\beta}_j}
		\sum_{l=1}^{n_{j}}
		(\sqrt{\gamma_i}) \veta^{(i)}_k \cdot \nabla_{x^{(i)}_k}
		\left(
			\veta^{(j)}_l \cdot \nabla_{x^{(j)}_l}
			g^{\vn}(\vx^{\vn}, t)
		\right)
	-  (\sqrt{\gamma_i}) \veta^{(i)}_k \cdot \nabla_{x^{(i)}_k}
	   \chi^{\vn} (\vx^{\vn}, t) \\
	&\quad=
		\sum_{j=1}^J
		\frac{\sqrt{\gamma_j \gamma_i}}{\hat{\beta}_j}
		\sum_{l=1}^{n_{j}}
		\veta^{(i)}_k  \veta^{(j)^T}_l
		::
		\nabla_{x^{(i)}_k} \nabla_{x^{(j)}_l}^T g^{\vn}
	-  (\sqrt{\gamma_i}) \veta^{(i)}_k \cdot \nabla_{x^{(i)}_k}
	   \chi^{\vn} (\vx^{\vn}, t),
\end{align*}
where, for two square matrices $A$ and $B \in \R^{n \times n}$,
the notation $A::B$ denotes the inner product on square matrices, i.e.
$A::B := tr(AB^T) = \sum_{i=1}^n \sum_{j=1}^n a_{ij} b_{ij}$.
Hence,
\begin{align*}
	&\hat{\Lp}^{(2)} f^{\vn}_1
	=
		\sum_{i=1}^J \sum_{k=1}^{n_i} \sqrt{\gamma_i}\brac{
		\sum_{j=1}^J \sum_{l=1}^{n_{j}}
		\tfrac{\sqrt{\gamma_j}}{\hat{\beta}_j}
		\Big(
		\veta^{(i)}_k  \veta^{(j)^T}_l
		\hspace{-0.5em} ::
		\nabla_{x^{(i)}_k} \nabla_{x^{(j)}_l}^T g^{\vn}
		\Big)
		-  \veta^{(i)}_k \nabla_{x^{(i)}_k}
	   \chi^{\vn} (\vx^{\vn}, t)}.
\end{align*}
We then have that
\begin{equation*}
	\int_{\R^{|\vn| d}}
	\paren{\hat{\Lp}^{(2)} f^{\vn}_1(\vzeta^{\vn}, t)}
	\rho^\infty(\veta^{\vn}) \, \mathrm{d} \veta^{\vn}
	= \sum_{j=1}^J D_j \Delta_{\vx^{n_j}} g^{\vn}(\vx^{\vn}, t),
\end{equation*}
by exploiting that the dropped terms of the integrand are odd functions.
The $O(1)$ solvability condition, \cref{s4 eq: sol cond O(1)}, then becomes
\begin{equation*}
\begin{aligned}
	&0 =
	\sum_{j=1}^J D_j \Delta_{\vx^{n_j}} g^{\vn}(\vx^{\vn}, t)
	- \frac{\partial g^{\vn}}{\partial t}
	-  \bigg(
		  \sum_{\vx^{\vn}_{\va} \in \Xip(\vx^{\vn})} K_{+} (\vx^{\vn}_{\va})
		  + \sum_{\vx^{\vn}_{\vb} \in \Xim(\vx^{\vn})} K_{-} (\vx^{\vn}_{\vb})
		\bigg) g^{\vn} \\
	&\qquad
		+ \Rpp^\dagger[g^{\vnp}] (\vx^{\vn}, t)
		+ \Rmm^\dagger[g^{\vnm}] (\vx^{\vn}, t),
\end{aligned}
\end{equation*}
representing the dynamics for the over-damped limit.

In summary, we find that the leading-order spatial densities
\begin{equation}
	\int_{\R^{|\vn| d}}
	p^{(\vn)}(\vxi^{\vn}, t) \mathrm{d} \vv^{\vn}
	\sim g^{\vn}(\vx^{\vn}, t), \quad (\beta \to \infty)
\end{equation}
 satisfy the standard over-damped volume reactivity PBSRD model (see
 \cite{IsaacsonZhang17,isaacson2022mean})
\begin{equation*}
\begin{aligned}
	\frac{\partial g^{\vn}}{\partial t}
	&=  \sum_{j=1}^J D_j \Delta_{\vx^{n_j}} g^{\vn}(\vx^{\vn}, t) \\
	&\qquad - \bigg(
		\sum_{\vx^{\vn}_{\va} \in \Xi_+(\vx^{\vn})} K_{+} (\vx^{\vn}_{\va})
	  + \sum_{\vx^{\vn}_{\vb} \in \Xi_-(\vx^{\vn})} K_{-} (\vx^{\vn}_{\vb})
	 \bigg)
	g^{\vn} (\vx^{\vn}, t) \\
	&\qquad +
		\sum_{\vx^{\vn}_{\vb} \in \Xim(\vx^{\vn})}
		\rev{\frac{1}{\va !}} \int_{\Omega^{|\va|}}
		m_{+} (\vx^{\vn}_{\vb} | \vx^{\vnp}_{\va})
		K_+ (\vx^{\vnp}_{\va})
		g^{\vnp}((\vx^{\vn} \backslash \vx^{\vn}_{\vb}) \cup \vx^{\vnp}_{\va}, t)
		\mathrm{d} \vx^{\vnp}_{\va} \\
	&\qquad +
		\sum_{\vx^{\vn}_{\va} \in \Xip(\vx^{\vn})}
		\rev{\frac{1}{\vb !}} \int_{\Omega^{|\vb|}}
		m_{-} (\vx^{\vn}_{\va} | \vx^{\vnm}_{\vb})
		K_{-} (\vx^{\vnm}_{\vb})
		g^{\vnm}((\vx^{\vn} \backslash \vx^{\vn}_{\va}) \cup \vx^{\vnm}_{\vb}, t)
		\mathrm{d} \vx^{\vnm}_{\vb}.
\end{aligned}
\end{equation*}

\section{Examples}\label{s5} We now illustrate how the forward and backward
reaction kernels for the examples of \cref{s2} were obtained by enforcing
consistency with detailed balance, present the overdamped limits for these
examples, and demonstrate the results are consistent with the general case
studied in the previous section.

\begin{example}[$\sA + \sB \rightleftharpoons \sC$]\label{s5 eg: ABC}
    Recall \cref{s2 eg: ABC}, in which particles move via the Langevin Dynamics \cref{s2 eq: Langevin Dynamics}
    and can undergo the reversible reaction $\sA + \sB \rightleftharpoons \sC$.
    In this context, $\bm{P}(t) =$ $\{p_{12}(\xi_1, \xi_2, t),$ $p_{3}(\xi_3, t)\}$,
    where $p_{12}(\xi_1, \xi_2, t)$ denotes the probability density
    the particles are unbound at time $t$, with the $\sA$ particle having state $\xi_1$ and the $\sB$ particle state $\xi_2$. $p_{3}(\xi_3, t)$ represents the probabilty the particles are in the bound state at $t$, with the $\sC$ particle having state $\xi_3$. $\bm{P}(t)$ satisfies
    \begin{equation}\label{s5 eq: AB-C Kolmogorov eq}
    \begin{aligned}
        \PD{p_{1 2}}{t}
        &= (\mathcal{L}_1 + \mathcal{L}_2) p_{1 2} - K_+^{\beta}(\xi_1,\xi_2) p_{1 2}
        + \int_{\Omega \times \R^{d}} \hspace{-1.0em} K_-^{\beta}(\xi_3) m_-^{\beta}(\xi_1,\xi_2 | \xi_3) p_3(\xi_3,t) \, \mathrm{d} \xi_3, \\
        \PD{p_3}{t}
        &= \mathcal{L}_3 p_3 -K_-^{\beta}(\xi_3) p_3
        + \int_{(\Omega \times \R^d)^2} \hspace{-1.2em} K_+^{\beta}(\xi_1,\xi_2) m_+^{\beta}(\xi_3 | \xi_1,\xi_2) p_{1 2}(\xi_1,\xi_2,t) \, \mathrm{d} \xi_1 \mathrm{d} \xi_2,
    \end{aligned}
    \end{equation}
    where $\Lp_i$ for $i = 1, 2$ are hypoelliptic transport operators defined
    analogously to \cref{s3 eq: L}. The Kolmogorov forward equation \cref{s5 eq:
    AB-C Kolmogorov eq} is simply a special case of \cref{s3 eq: Kolmogorov forward
    eq}.

    Similar to the general case, we expect that the principle of detailed
    balance of pointwise reaction fluxes,
    \begin{equation}  \label{s5 eq: AB-C detailed balance}
        K_+^{\beta}(\xi_1,\xi_2) m_+^{\beta}(\xi_3 | \xi_1, \xi_2) \bar{p}_{1 2}(\xi_1, \xi_2)
        = K_-^{\beta}(\xi_3) m_-^{\beta}(\xi_1, \xi_2 | \xi_3) \bar{p}_{3}(\xi_3),
    \end{equation}
    should hold for the equilibrium solutions $\bar{p}_{12}(\xi_1, \xi_2)$ and $\bar{p}_{3}(\xi_3)$.
    Substituting into the steady-state equation for \cref{s5 eq: AB-C Kolmogorov eq}, this
    implies
     \begin{equation*}
        \bar{p}_{1 2}(\xi_1,\xi_2) = \frac{\pi_{1 2}}{\abs{\Omega}^2} \prod_{i = 1}^2 \GaussianDensity_{d}(v_i;(D_i \beta_i)\mathrm{I}_{d}),
        \qquad
        \bar{p}_3(\xi_3) = \frac{\pi_{3}}{\abs{\Omega}}
        \GaussianDensity_{d}(v_3;(D_3 \beta_3)\mathrm{I}_{d}),
    \end{equation*}
    where $\pi_{12}$ and $\pi_{3}$ denote the equilibrium probabilities to be
    in the \rev{unbound (i.e. $(A,B)$) vs. bound (i.e. $C$)} states. We assume these probabilities should be the
    same as in a standard well-mixed equilibrium model for the
    reaction, so that $\tfrac{\pi_{1
    2}}{\pi_{3}} = \Kd \abs{\Omega}$, where $\Kd$ denotes the dissociation
    constant of the reaction~\cite{ZhangIsaacson2022,van1976equilibrium}.

    By substituting the corresponding rate functions $K_{\pm}$ and forward
    placement densities $m_+^{\beta}$ of Tables \ref{s1 tab: rate
    functions}--\ref{s1 tab: v placement} into the detailed balance relation
    \cref{s5 eq: AB-C detailed balance}, we find the backward placement density
    must be given by
    \begin{equation} \label{eq:mminus_a_b_to_c_v1}
		\begin{aligned}
        m_-^{\beta}(\xi_1, \xi_2 | \xi_3) &= m_-(x_1, x_2 | x_3) \, m_-^{\beta}(v_1, v_2 | v_3) \\
        &= \tfrac{1}{\abs{B_\varepsilon}} \ind_{[0,\varepsilon]}(\abs{x_1-x_2})
        \delta(x_3 - (\alpha x_1 + (1-\alpha) x_2)) \\
        &\phantom{=}  \times \delta\paren{v_3 - \tfrac{(m_1 v_1 + m_2 v_2)}{m_3}} \GaussianDensity^{-1}_{d}(v_3;(D_3\beta_{3})\mathrm{I}_{d})
         \prod_{i = 1}^2 \GaussianDensity_{d}(v_i;(D_i\beta_{i})\mathrm{I}_{d}).
    \end{aligned}
	\end{equation}
    Here we have assumed that $\lambda_- := \Kd \lambda_+ \abs{B_\varepsilon}$
    (consistent with the detailed balance conditions for the overdamped case,
    see~\cite{ZhangIsaacson2022}). Note that $m_-(x_1, x_2 | x_3)$ is also the normalized
    spatial placement density of the corresponding over-damped model~\cite{ZhangIsaacson2022}.
		
    Assuming conservation of mass, i.e. $m_1 + m_2 = m_3$, and using the
    identities in \cref{eq:a_b_to_c_d_m_gamma_idents}, we see that
    $m_-^{\beta}(v_1, v_2 | v_3)$ is normalized. However, as written it is not
    clear what physical placement model it represents. Again applying the
    identities in \cref{eq:a_b_to_c_d_m_gamma_idents}, and using that the
    $\delta$-function determines the value of $v_3$, we find
    \begin{align}
	    m_-^{\beta}(v_1, v_2 | v_3) &= \paren{\tfrac{D_3 \beta_3}{2 \pi D_1 \beta_1 D_2 \beta_2}}^{d/2}
	    \!\!\!\delta\paren{v_3 - \tfrac{(m_1 v_1 + m_2 v_2)}{m_3}}
	    e^{\abs{v_3}^2 / 2 D_3 \beta_3} \prod_{i = 1}^2 e^{-\abs{v_i}^2 / 2 D_i \beta_i} \notag \\
	    &= \paren{\tfrac{1}{2\pi (D_1 \beta_1 + D_2 \beta_2)}}^{d/2}
	    \delta\paren{v_3 - \tfrac{(m_1 v_1 + m_2 v_2)}{m_3}}
	    e^{-\abs{v_1 - v_2}^2/(2(D_1 \beta_1 + D_2 \beta_2))}. \label{eq:unbind_velo_kernel_final}
    \end{align}
	\cref{eq:unbind_velo_kernel_final} can then be interpreted as enforcing that
    total momentum is conserved in the unbinding reaction, and that the
    separation velocity of the products satisfies a Maxwell-Boltzmann
    distribution (i.e. that $v_1 - v_2 \sim
    \mathcal{N}(0,(D_1\beta_1+D_2\beta_2)I_d)$). This is consistent with the
    form we gave in \cref{s2 eg: ABC}.

    Finally, we now sketch the direct overdamped limit of \cref{s5 eq: AB-C
    Kolmogorov eq}, and show it is consistent with the general result of the
    last section. Consider the factorization
    \begin{align*}
        p_{12}(\xi_1, \xi_2, t) := \rho_{12}(\xi_1, \xi_2, t) \bar{u}_{12}, &&
        p_{3}(\xi_3, t) := \rho_{3}(\xi_3, t) \bar{u}_{3},
    \end{align*}
    where
    \begin{align*}
        \bar{u}_{1 2}(\xi_1,\xi_2)  = \prod_{i = 1}^2  \GaussianDensity_{d}(v_i;(D_i\beta_{i})\mathrm{I}_{d}),
        && \bar{u}_{3}(\xi_3) =  \GaussianDensity_{d}(v_3;(D_3\beta_{3})\mathrm{I}_{d}).
    \end{align*}
    We first substitute the above factorization into the Kolmogorov equation \cref{s5 eq: AB-C Kolmogorov eq},
    which gives the forward equations that $\rho_{12}$ and $\rho_{3}$ satisfy similarly to \cref{s4 eq: Kolmogorov rho}.
    Then, we rewrite the transport operator $\Lp$ and the velocity placement kernels
    $m^{\beta}_+(v_3|v_1, v_2)$ and $m^{\beta}_-(v_1, v_2|v_3)$ under the new coordinates
    $v_i = \sqrt{\beta_i D_i} \eta_i = \sqrt{\beta \gamma_i} \eta_i$.
    By defining $\zeta_i = (x_i, \eta_i)$,
    $f_{12}(\zeta_1, \zeta_2, t) := \rho_{12}(\xi_1, \xi_2, t)$, and
    $f_{3}(\zeta_3, t) := \rho_{3}(\xi_3, t)$,
    we get the forward equation for $f_{12}$ and $f_{3}$ as follows
    \begin{equation}\label{s5 eq: AB-C forward f}
    \begin{aligned}
        \PD{f_{1 2}}{t} &= \sum_{i=1}^2 \paren{\beta \hat{L}^{(1)}_i + \sqrt{\beta}\hat{L}^{(2)}_i} f_{1 2} - K_+(x_1,x_2) f_{1 2} + \mathcal{R}_-\brac{f_3}(\zeta_1,\zeta_2,t) \\
        \PD{f_{3}}{t} &= \paren{\beta \hat{L}^{(1)}_3 + \sqrt{\beta} \hat{L}^{(2)}_3} f_{3} - K_-(x_3) f_{3} + \mathcal{R}_+\brac{f_{1 2}}(\zeta_3,t).
    \end{aligned}
    \end{equation}
    where $\hat{L}^{(1)}_i$ and $\hat{L}^{(2)}_i$ are the non-dimensionalized
    transport operators
	\begin{equation*} \hat{L}_i^{(1)} = \hat{\beta_i}
        \paren{\Delta_{\eta_i} - \eta_i \cdot \nabla_{\eta_i}},
        \qquad
        \hat{L}_i^{(2)} = - \paren{\sqrt{\gamma_i}} \eta_i \cdot \nabla_{x_i},
    \end{equation*}
    and
    \begin{align*}
        \mathcal{R}_+\brac{f_{12}}(\zeta_3,t)
        &=
        \tfrac{1}{(2 \pi)^{d/2}}\int_{(\Omega \times \R^d)^2} K_+(x_1,x_2) m_{+}(x_3 | x_1,x_2)  f_{1 2}(\zeta_1,\zeta_2,t) \\
	    & \hspace{-4.0em} \phantom{=} \qquad\qquad  \times
        \delta\!\paren{\eta_3 - \paren{\sqrt{\tfrac{\gamma_3}{\gamma_1}}\eta_1 +  \sqrt{\tfrac{\gamma_3}{\gamma_2}}\eta_2}}
        e^{-\abs{\sqrt{\gamma_1}\eta_1 - \sqrt{\gamma_2} \eta_2}^2 / 2(\gamma_1 + \gamma_2)} \,
        \mathrm{d} \zeta_1 \mathrm{d} \zeta_2 , \\
        \mathcal{R}_-\brac{f_{3}}(\zeta_1, \zeta_2, t)
        &=
        \int_{\Omega \times \R^d} \hspace{-1.5em} K_-(x_3) m_{-}(x_1,x_2 | x_3)  f_3(\zeta_3, t) \delta\!\paren{\eta_3 - \paren{\sqrt{\tfrac{\gamma_3}{\gamma_1}}\eta_1 +  \sqrt{\tfrac{\gamma_3}{\gamma_2}}\eta_2}} \, \mathrm{d} \zeta_3  .
    \end{align*}

    We now develop the asymptotic expansion of $f_{12}$ and $f_3$ as $\beta \rightarrow \infty$
    of the form
    \begin{align*}
        f_{1 2}(\zeta_1,\zeta_2,t) &\sim \fab{0}(\zeta_1,\zeta_2,t) + \tfrac{1}{\sqrt{\beta}} \fab{1}(\zeta_1,\zeta_2,t) + \tfrac{1}{\beta} \fab{2}(\zeta_1,\zeta_2,t) + \dots, \\
        f_{3}(\zeta_3,t) &\sim \fc{0}(\zeta_3,t) + \tfrac{1}{\sqrt{\beta}} \fc{1}(\zeta_3,t) + \tfrac{1}{\beta} \fc{2}(\zeta_3,t) + \dots.
    \end{align*}
    Similar to what we did in \cref{s4: overdamped limit}, we substitute the
    expansions of $f_{12}$ and $f_3$ into the forward equations \cref{s5 eq:
    AB-C forward f} respectively, and balance the terms based on the different
    orders of the friction constant $\beta$ as we did in (\ref{s4 eq: O
    beta})-(\ref{s4 eq: O 1}). From this point on the analysis is similar to
     \cref{s4: overdamped limit}, yielding the standard two-particle
    Volume-Reactivity PBSRD model for $\sA + \sB \leftrightarrows \sC$
    (see~\cite{IsaacsonZhang17,ZhangIsaacson2022}). That is, as $\beta \to \infty$
	\begin{align*}
		\int_{\R^{2d}} p_{12}(\xi_1,\xi_2,t) \, \mathrm{d}v_1 \mathrm{d} v_2  \sim g_{12}(x_1,x_2,t)
	&&\text{and}
	&&\int_{\R^{d}} p_{3}(\xi_3,t) \, \mathrm{d} v_3 \sim g_3(x_3,t),
	\end{align*}
	where $g_{12}$ and $g_3$ satisfy the two-particle VR PBSRD model
    \begin{align*}
        \PD{g_{1 2}}{t} &= (D_1 \lap_{x_1} \!+ D_2 \lap_{x_2}) g_{1 2} - K_+(x_1,x_2) g_{1 2}
        + \!\! \int_{\Omega} \!\! K_-(x_3) m_-(x_1,x_2 | x_3) g_3(x_3,t) \, \mathrm{d} x_3, \\
        \PD{g_{3}}{t} &= D_3 \lap_{x_3} g_{3} - K_-(x_3) g_{3} + \int_{\Omega^2} K_+(x_1,x_2) m_+(x_3 | x_1, x_2) g_{1 2}(x_1, x_2, t) \, \mathrm{d} x_1 \mathrm{d} x_2.
    \end{align*}
\end{example}
\begin{example}[$\sA + \sB \rightleftharpoons \sC + \sD$]\label{s5_ex_AB-CD} We
    next consider the two-particle system undergoing the Langevin Dynamics \cref{s2 eq:
    Langevin Dynamics} with reversible reaction $\sA + \sB \rightleftharpoons
    \sC + \sD$. In this context, $\bm{P}(t) = \{p_{12}(\xi_1, \xi_2, t),
    p_{34}(\xi_3, \xi_4, t)\}$, and satisfies
    \begin{equation}\label{s5 eq: AB-CD Kolmogorov eq}
    \begin{aligned}
        \PD{p_{12}}{t} &= (\mathcal{L}_1 + \mathcal{L}_2) p_{1 2}  - K_+^{\beta}(\xi_1,\xi_2) p_{1 2}(\xi_1,\xi_2,t) \\
        &\qquad+ \int_{(\Omega \times \R^d)^2} m_-^{\beta}(\xi_1,\xi_2 | \xi_3, \xi_4) K_-^{\beta}(\xi_3,\xi_4) p_{3 4}(\xi_3,\xi_4,t) \, \mathrm{d}\xi_3 \mathrm{d}\xi_4, \\
        \PD{p_{34}}{t} &= (\mathcal{L}_3 + \mathcal{L}_4) p_{3 4} -K_-^{\beta}(\xi_3,\xi_4) p_{3 4}(\xi_3,\xi_4,t)\\
        &\qquad+ \int_{(\Omega \times \R^d)^2} m_+^{\beta}(\xi_3, \xi_4 | \xi_1,\xi_2) K_+^{\beta}(\xi_1,\xi_2) p_{1 2}(\xi_1,\xi_2,t) \, \mathrm{d}\xi_1 \mathrm{d}\xi_2,
    \end{aligned}
    \end{equation}
    where each $\Lp_i$ is a hypoelliptic transport operator defined analogously
     to \cref{s3 eq: Lp} from the general case. The Kolmogorov forward equation
     \cref{s5 eq: AB-CD Kolmogorov eq} is a special case of \cref{s3 eq:
     Kolmogorov forward eq}.

    Similar to the general case, we assume the principle of detailed balance of
    pointwise reaction fluxes,
    \begin{equation} \label{s5 eq: AB-CD detailed balance}
        m_+^{\beta}(\xi_3,\xi_4 | \xi_1, \xi_2) K_+^{\beta}(\xi_1,\xi_2)  \bar{p}_{1 2}(\xi_1, \xi_2)
        = m_-^{\beta}(\xi_1, \xi_2 | \xi_3,\xi_4) K_-^{\beta}(\xi_3, \xi_4) \bar{p}_{3 4}(\xi_3, \xi_4),
    \end{equation}
    holds for the equilibrium solutions $\bar{p}_{1 2}(\xi_1,\xi_2)$ and $\bar{p}_{3 4}(\xi_3,\xi_4)$.
    Substituting \cref{s5 eq: AB-CD detailed balance} into (\ref{s5 eq: AB-CD Kolmogorov eq}) gives that
    \begin{align*}
        \bar{p}_{1 2}(\xi_1,\xi_2) &= \tfrac{\pi_{1 2}}{\abs{\Omega}^2} \bar{p}_{1 2}(v_1,v_2), & &\text{where} &
		\bar{p}_{1 2}(v_1,v_2) &:= \prod_{i = 1}^2 \GaussianDensity_{d}(v_i;(D_i \beta_i)\mathrm{I}_{d}), \\
        \bar{p}_{3 4}(\xi_3,\xi_4)&= \tfrac{\pi_{3 4}}{\abs{\Omega}^{2}} \bar{p}_{3 4}(v_3,v_4), & &\text{where} &
		 \bar{p}_{3 4}(v_3,v_4) &:= \prod_{i = 3}^4 \GaussianDensity_{d}(v_i;(D_i \beta_i)\mathrm{I}_{d}).
    \end{align*}
    Here, $\pi_{12}$ and $\pi_{34}$ denote the equilibrium probabilities to be
    in the unbound vs. bound states. As in the last example, we assume they
    should be consistent with the corresponding well-mixed chemical master
    equation equilibrium model for the reaction, so that $\tfrac{\pi_{1
    2}}{\pi_{3 4}} = \Kd$, where $\Kd$ denotes the dissociation constant of the
    reaction.

	Let $\mathrm{p}_{i j} := m_i v_i + m_j v_j$, the total mass be $\bar{m} := m_1
    + m_2 = m_3 + m_4$, and assume that $\lambda_- := \Kd \lambda_+$ (again
    consistent with the detailed balance conditions for the overdamped case). By
    substituting the corresponding rate functions $K_{\pm}$ and forward
    placement densities $m_+^{\beta}$ of Tables \ref{s1 tab: rate
    functions}--\ref{s1 tab: v placement} into the detailed balance relation
    \cref{s5 eq: AB-CD detailed balance}, we find the backward placement density
    must be given by $m_-^{\beta}(\xi_1, \xi_2 | \xi_3, \xi_4) = m_-(x_1, x_2 |
    x_3, x_4) \, m_-^{\beta}(v_1, v_2 | v_3, v_4)$, where $m_-(x_1,x_2 | x_3,
    x_4)$ is given by \cref{s1 tab: x placement} and
	\begin{equation} \label{eq:m_minus_ab_cd_nonsimple}
        m_-^{\beta}(v_1, v_2 | v_3, v_4) =
		\Big[ \tfrac{\bar{p}(v_1,v_2)}{\bar{p}(v_3,v_4)}
		 \bar{m}^d \delta\paren{\mathrm{p}_{3 4} - \mathrm{p}_{1 2}}
		\GaussianDensity_{d}(v_3-v_4;(D_3\beta_3+D_4\beta_4)\mathrm{I}_{d}) \Big].
	\end{equation}


    We now confirm this reduces to the formula in \cref{s1 tab: v placement},
    and is properly normalized. Showing that the forward velocity placement
    density is also normalized follows by a similar calculation. In the context
    of \cref{eq:m_minus_ab_cd_nonsimple}, using the Einstein relations,
    \cref{eq:a_b_to_c_d_param_ident_2}, and conservation of mass, we have that
	\begin{multline} \label{eq:mb_idents}
		\frac{\GaussianDensity_{d}(v_3-v_4;(D_3\beta_3+D_4\beta_4)\mathrm{I}_{d})}{\bar{p}(v_3,v_4)}
		= \frac{1}{\bar{m}^d \GaussianDensity_{d}(\mathrm{p}_{3 4};(\kb T \bar{m})\mathrm{I}_{d})} \\
		= \frac{1}{\bar{m}^d \GaussianDensity_{d}(\mathrm{p}_{1 2};(\kb T \bar{m})\mathrm{I}_{d})}
		= \frac{\GaussianDensity_{d}(v_1-v_2;(D_1\beta_1+D_2\beta_2)\mathrm{I}_{d})}{\bar{p}(v_1,v_2)},
	\end{multline}
	where, in the second line, we also used that the $\delta$-function sets
	$\mathrm{p}_{1 2} = \mathrm{p}_{3 4}$. Substituting into
	\cref{eq:m_minus_ab_cd_nonsimple} gives the formula in \cref{s1 tab: v
	placement}. When a $\sC + \sD \to \sA + \sB$ reaction occurs, the formula
	corresponds to sampling the velocities of two product particle such that the
	total product momentum equals the total substrate momentum, and the
	products' relative velocity is sampled from a Maxwell-Boltzmann
	distribution.

	To confirm the normalization note that
    \begin{multline*}
	    \int_{\R^{2d}}\bar{p}_{1 2}(v_1,v_2) \delta\paren{\mathrm{p}_{3 4} - \mathrm{p}_{1 2}} \, \mathrm{d} v_1 \mathrm{d} v_2
		= \GaussianDensity_{d}(\mathrm{p}_{3 4};(D_1\beta_1 m_1^{2}+D_2\beta_2 m_2^{2})\mathrm{I}_{d}) \\
		= \GaussianDensity_{d}(\mathrm{p}_{3 4};(D_3\beta_3 m_3^{2}+D_4\beta_4 m_4^{2})\mathrm{I}_{d})
		= \GaussianDensity_{d}(\mathrm{p}_{3 4};(\kb T \bar{m})\mathrm{I}_{d}),
    \end{multline*}
	where we have used the identies \cref{eq:a_b_to_c_d_param_ident_1} and
	\cref{eq:a_b_to_c_d_param_ident_2}. Combining with the first identity in
	\cref{eq:mb_idents}, we see that $m_-^{\beta}$ is normalized in $(v_1,v_2)$.

    From this point on, the analysis is similar to \cref{s5 eg: ABC}.
    We find that in the over-damped limit $\beta \to \infty$,
	\begin{equation*}
		\int_{\R^{2d}} p_{i j}(\xi_i, \xi_j,t) \, \mathrm{d}v_i \mathrm{d}v_j \sim g_{i j}(x_i, x_j,t),
		\quad (i,j) \in \{(1,2),(3,4)\},
	\end{equation*}
	where $g_{i j}(x_i,x_j,t)$ satisfy the Doi VR PBSRD model
    \begin{align*}
        \PD{g_{1 2}}{t} &= (D_1 \lap_{x_1} + D_2 \lap_{x_2}) g_{1 2}  - K_+(x_1,x_2) g_{1 2} \\
        &\qquad
        + \int_{\Omega^2} K_-(x_3,x_4) m_-(x_1,x_2 | x_3,x_4) g_{3 4}(x_3,x_4,t) \, \mathrm{d}x_3 \mathrm{d}x_4, \\
        \PD{g_{3 4}}{t} &= (D_3 \lap_{x_3} + D_4 \lap_{x_4}) g_{3 4} - K_-(x_3,x_4) g_{3 4} \\
        &\qquad+ \int_{\Omega^2} K_+(x_1,x_2) m_+(x_3, x_4 | x_1, x_2) g_{1 2}(x_1, x_2, t) \, \mathrm{d}x_1 \mathrm{d}x_2.
    \end{align*}
\end{example}
\begin{remark}
	\rev{In the preceding examples, we provided explicit formulas for the
	reactive rate functions and displacement densities corresponding to several
	common, reversible reactions. Reactive kernels for other reactions can be
	defined analogously. Our general theory still holds, as shown in the
	previous section, for reversible reactions of arbitrary order. For readers
	interested in modelling such higher-order reactions, i.e. greater than order
	two, one could define appropriate higher-order reactive interaction kernels
	and then apply our theory. For example, in the overdamped case, kernels for
	Doi PBSRD reaction models that are higher than second order have been
	proposed in~\cite{FleggReversibleNthOrder2019}, and could be modified in
	analogous manner to our preceding examples to derive kernels for use in LD
	models. Alternatively, one could follow the approach of Plesa
	\cite{plesa2023stochastic} and approximate higher-order reactions via a
	series of first- and second-order reactions.}
\end{remark}

\section{Numerical Simulation}\label{s6}

\begin{figure}
    \centering
    \begin{minipage}{0.5\textwidth}
        \centering
        \includegraphics[width=0.99\textwidth]{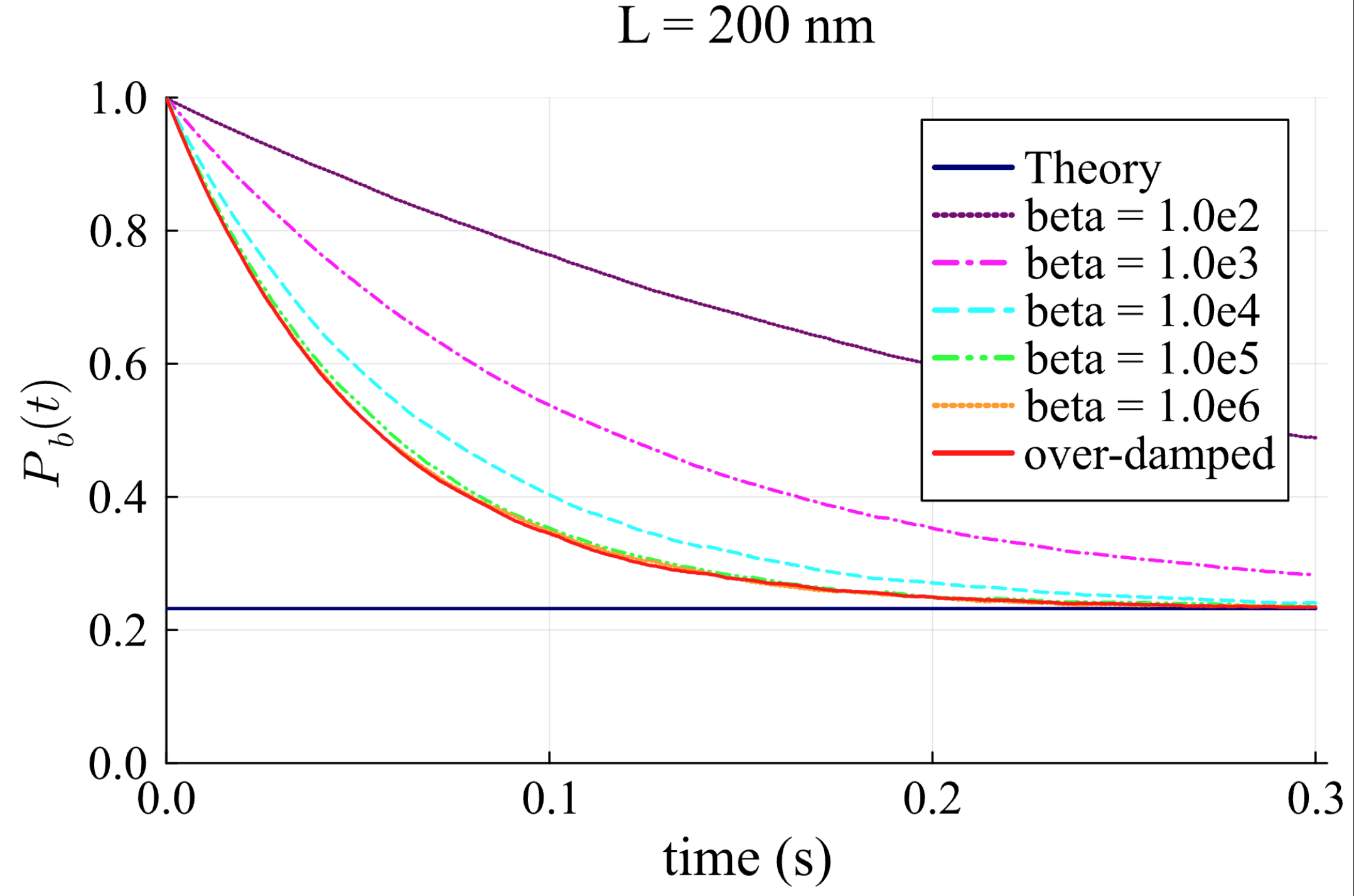} 
    \end{minipage}
    \begin{minipage}{0.5\textwidth}
        \centering
        \includegraphics[width=0.99\textwidth]{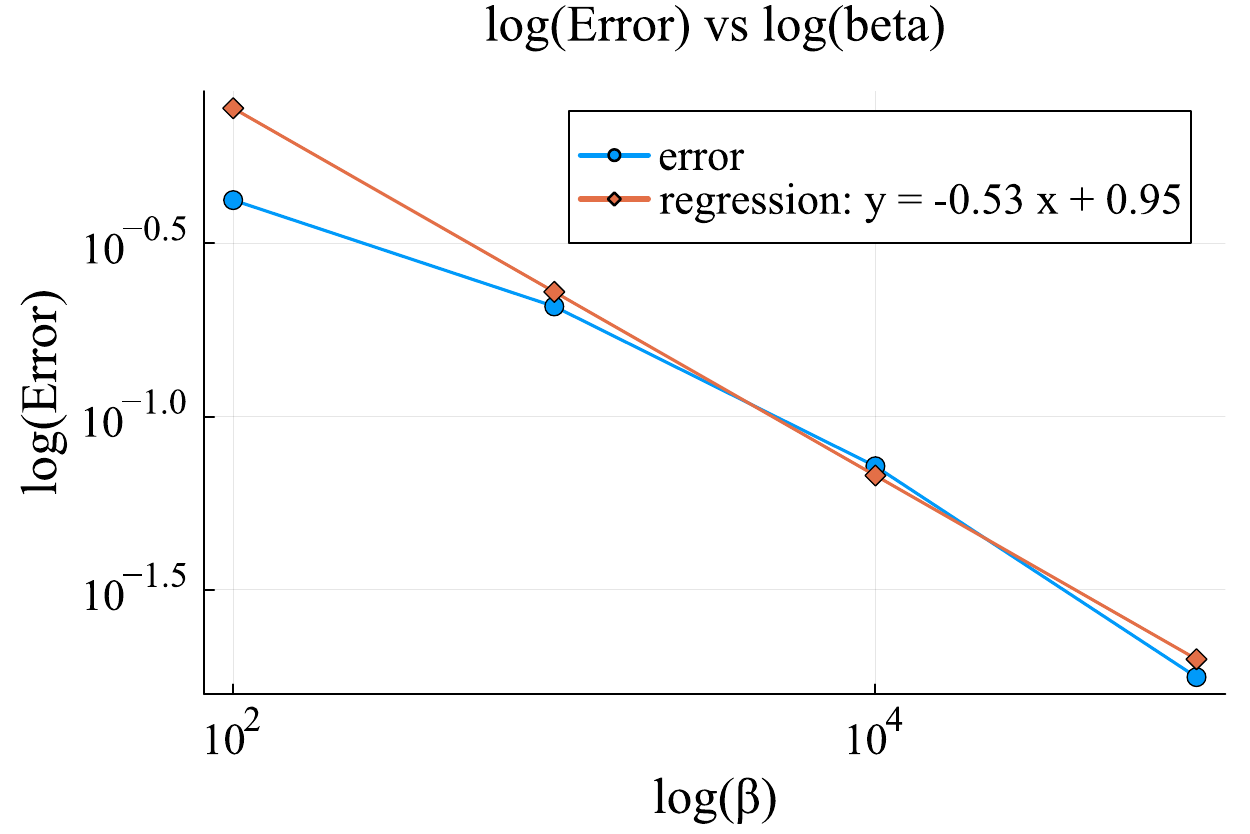} 
    \end{minipage}
    \caption{(left) Convergence of reactive Langevin Dynamics (RLD) to
            over-damped reactive Brownian Dynamics (RBD) as the friction
            constant $\beta$ (in units of $\text{s}^{-1}$) increases for the
            two-particle $\sA + \sB \leftrightarrow \sC$ reaction. \rev{The left
            panel shows the probabilty the two particles are in the bound, i.e.
            $C$, state as a function of time, denoted by $P_b(t)$. It also}
            illustrates convergence as $t \to \infty$ \rev{of $P_b(t)$ in both} models
            to the steady-state of the corresponding well-mixed chemical master
            equation model ("Theory"). (right) Maximum difference between RLD
            and RBD estimates for $P_b(t)$ as $\beta$ is increased. \rev{The
            regression slope, -0.53, is consistent with the scaling,
            $1/\sqrt{\beta}$, of the first omitted term of the expansion
            (\ref{s4 eq: expansion})}. }
    \label{fig}
\end{figure}

To illustrate the asymptotic behavior as $\beta \to \infty$ of the reactive
Langevin Dynamics model derived in the previous sections, we numerically studied
a RLD model for the reversible reaction $\sA + \sB \rightleftharpoons \sC$ in
the special case of a system with just one $\sC$ particle at $t = 0$. We
successively increased the friction constant $\beta$ to demonstrate that the
empirical overdamped limit of the RLD model is consistent with the corresponding
VR PBSRD model.

\begin{algorithm}[htpb]
	\caption{Numerical method for simulating RLD model of $\sA + \sB \rightleftharpoons \sC$.}
	\label{alg: pseudo-code}
	\begin{algorithmic}[1]
		\STATE \textbf{function} periodic($X$; $L$)
			\STATE \quad \textbf{return} $X = [ \text{mod} (x, L) \text{ for } x \text{ in } X]$
		\STATE
		\STATE \textbf{function} implicit\_euler($X_t$, $V_t$;\, $\Delta t$, $\beta$, $D$)
			\STATE \quad generate $Z \sim \mathcal{N}(0, \mathrm{I}_3)$
			\STATE \quad $V_{t+\Delta t} = (V_t + \beta \sqrt{2 D \Delta t} Z) / (1 + \beta \Delta t)$ 
			\STATE \quad $X_{t+\Delta t} = \text{periodic} \big(X_t + V_{t+\Delta t} \Delta t; L\big)$ 
			\STATE \quad \textbf{return} $(X_{t+\Delta t}, V_{t+\Delta t})$
		\STATE
		\STATE{Initialize $X^{C}_0 \sim \mathcal{U}([0,L]^3)$ and $V^{C}_0 \sim \mathcal{U}([-1.0e7, 1.0e7]^3)$}
		\STATE{\texttt{state} = 0 (system contains single particle $\sC$)}
		\FOR{$i = 1,..., \lfloor T / \Delta t \rfloor$}
			\IF{\texttt{state} = 0}
				
				\STATE{$ X^{C}_{(i+1)\Delta t}, V^{C}_{(i+1)\Delta t} = \text{implicit\_euler} (X^{C}_{i\Delta t}, V^{C}_{i\Delta t};\, \Delta t, \beta, D$)}
				\IF{$U[0,1] \leq \lambda_- \Delta t$}
					\STATE{generate $\eta_i \sim U(B(0, \varepsilon))$}
					\STATE{place $X^{A}_{(i+1) \Delta t}$ and $X^{B}_{(i+1) \Delta t}$ by solving the following linear system}
						\STATE{$\qquad \alpha  X^{A}_{(i+1)\Delta t} + (1-\alpha) X^{B}_{(i+1)\Delta t} = X^{C}_{i\Delta t}$}
						\STATE{$\qquad X^{A}_{(i+1)\Delta t} - X^{B}_{(i+1)\Delta t} = \eta_i$}
					\STATE{$X^{A}_{(i+1)\Delta t} = \text{periodic}(X^{A}_{(i+1) \Delta t}; L)$, $X^{B}_{(i+1) \Delta t} = \text{periodic}(X^{B}_{(i+1) \Delta t}; L)$}
					\STATE{generate $\zeta_i \sim \mathcal{N}(0, (D_1 \hat{\beta}_1 \beta + D_2 \hat{\beta}_2 \beta) \mathrm{I}_3)$}
					\STATE{place $V^{A}_{(i+1) \Delta t}$ and $V^{B}_{(i+1) \Delta t}$ by solving the following linear system}
						\STATE{$\qquad \frac{m_1}{m_3} V^{A}_{(i+1)\Delta t} + \frac{m_2}{m_3} V^{B}_{(i+1)\Delta t} = V^{C}_{i\Delta t}$}
						\STATE{$\qquad V^{A}_{(i+1)\Delta t} - V^{B}_{(i+1)\Delta t} = \zeta_i$}
					\STATE{\texttt{state} = 1}
				\ENDIF
			\ELSE
					\STATE{$X^{A}_{(i+1)\Delta t}, V^{A}_{(i+1)\Delta t} = \text{implicit\_euler} (X^{A}_{i\Delta t}, V^{A}_{i\Delta t};\, \Delta t, \beta, D$)}
					\STATE{$X^{B}_{(i+1)\Delta t}, V^{B}_{(i+1)\Delta t} = \text{implicit\_euler} (X^{B}_{i\Delta t}, V^{B}_{i\Delta t};\, \Delta t, \beta, D$)}
					\IF{$\texttt{periodic\_distance}(X^{A}_{(i+1)\Delta t} - X^{B}_{(i+1)\Delta t}) \leq \varepsilon$
						and $U[0,1] \leq \lambda_+ \Delta t$}
						\STATE{$X^{C}_{(i+1)\Delta t} = \text{periodic}(\alpha X^{A}_{i\Delta t} + (1-\alpha) X^{B}_{i\Delta t}; L)$}
						\STATE{$V^{C}_{(i+1)\Delta t} = \frac{m_1}{m_3} X^{A}_{i\Delta t} + \frac{m_2}{m_3} X^{B}_{i\Delta t}$}
						\STATE{\texttt{state} = 0}
					\ENDIF
			\ENDIF
			\STATE{save the \texttt{state} at $i$-th step in the current simulation path}
		\ENDFOR
	\end{algorithmic}
\end{algorithm}

\renewcommand{\arraystretch}{1.2}
\begin{table}[tbp]
	\small
    \caption{Parameters for Simulations}
    \centering
    \begin{tabular}{c | c r l}
    \hline\hline 
    Parameter & Value & Unit & Description \\ [0.5ex] 
    \hline
    $L$             & 200       & nm
    & domain length \\
    $(T,\Delta t)$             & (0.3, 1.0e-6)       & (s,s)
    & (final time, time step size)    \\
    $(\lambda_+,\lambda_-)$     & (1.0e4, 17.3)     &$(\text{s}^{-1},\text{s}^{-1})$
    & (association rate, dissociation rate)\\
    $\varepsilon$   & 10         & nm
    & reaction radius \\
    $\alpha$      & 0.5      & ratio
    & forward/backward placement ratio \\
    $D_i,\, i \in \{1,2,3\}$             & 1.0e6      & nm
    & diffusion coefficient \\
    $\hat{\beta}_1, \hat{\beta}_2$ & 1.0      & -
    & friction constant factor for $\sA$ and $\sB$\\
    [1ex] 
    \hline
    \end{tabular}
    \label{table: parameter}
\end{table}
	
We considered dynamics within a cubic domain, $\Omega = [0,L]^3$. In each
simulation, one $\sC$ particle was initially placed using a uniform spatial
density over $\Omega$, with initial velocity sampled from a uniform distribution
$\mathcal{U}([-1.0e7, 1.0e7]^3)$ nm/s. This was chosen to avoid particles
starting at equlibrium (\ref{s3 eq: limit density}). Spatial boundaries were
treated as periodic. \rev{Reactive interaction kernels therefore used periodic
distances in place of Euclidean distances to determine if particles were close
enough to react, and to determine where to place reaction product particles.} Our
Langevin-dynamics based algorithm is presented in \cref{alg: pseudo-code}, and
uses a fixed-timestep implicit Euler method to solve the SDEs for particle
transport. The parameters we used in simulations are given in \cref{table:
parameter}. Our reactive Brownian Dynamics method for the overdamped case was
the same we used in~\cite{ZhangIsaacson2022}.

To investigate the asymptotic behavior as $\beta \to \infty$, we varied $\beta
\in \{10^i \, (\text{s}^{-1})\}_{i=2}^6$. For each $\beta$, we performed
$N=50,000$ simulations and calculated the fraction of simulations in which the
system contained one $\sC$ particle at time $t$. This provided an empirical
estimate for $P_b(t)$, the probability the system was in the bound state at $t$.

In Figure \ref{fig} (left), we show $P_b(t)$ as $\beta$ is varied, along with
the over-damped limt from direct simulation of the corresponding VR PBSRD model.
As $\beta$ increases, we see that solutions to the RLD model converge to the
overdamped solution, which is consistent with our asymptotic analysis of the
preceding sections. In addition, we show that all solutions converge as $t \to
\infty$ to the equlibrium value for the analogous well-mixed chemical master
equation model, $\bar{P}_b = 1/(1 + \Kd \abs{\Omega})$ ("Theory" curve),
see~\cite{ZhangIsaacson2022}. In this specific instance, $\bar{P}_b = 0.2323$.
Figure \ref{fig} (right) displays the maximum difference across all timesteps of
$P_b(t)$ from each RLD model to the overdamped limit for varying $\beta$-values,
which further illustrates convergence as $\beta \to \infty$. \rev{Figure
\ref{fig} also highlights the significant quantitative differences between the
underdamped (Langevin dynamics with sufficiently small $\beta$) and overdamped
(Brownian dynamics) cases. This serves as an important reminder that the choice
between Brownian and Langevin dynamics should be guided by the biology and
physics of the problem one is studying. An inappropriate choice may lead to
models that fail to capture the correct physical behavior and yield inaccurate
results.}


\section{Conclusions}\label{s7} In this work, assuming the Einstein relation,
 assuming conservation of momentum and mass in reactions, and enforcing
 consistency with pointwise detailed balance of reactive fluxes at equilibrium,
 we formulated reactive interaction kernels for particle-based reactive Langevin
 dynamics (RLD) models of reversible reactions. For general reversible
 reactions, we then showed via asymptotic expansions that in the overdamped
 limit the derived kernels result in the RLD model converging to the classical
 volume reactivity particle-based stochastic reaction diffusion (PBSRD) model.
 In this way, our work provides a step towards, and illustrates contraints in,
 developing microscopic reactive Langevin-Dynamics models that remain fully
 consistent with widely-used overdamped reaction-diffusion models.

 There are a number of interesting followup questions that could be explored. It
 would be of mathematical interest to rigorously prove the overdamped limit,
 which is well-established in the absence of reactions. The presence of
 reactions is expected to complicate the mathematical analysis in potentially
 interesting ways.

 \rev{It is also clear from our analysis that more general forms of the reaction
 kernels could be assumed as long as their leading order behavior as $\beta \to
 \infty$ matches the behavior (i.e. $\beta$ scaling) assumed in this work.} In
 this way one could potentially relax the assumptions of conservation of mass or
 momentum that we made, and/or consider kernels that more closely model a
 specific microscopic reaction process (which may not be separable in $x$ and
 $v$).

 We have also assumed a relatively simple mass/friction model, which could be
 made substantially more realistic for specific biological applications. \rev{
 Similarly, the results of Erban suggest that a simple Langevin dynamics
 model as we study could fail to accurately describe ion transport in aqueous
 solutions \cite{erban2016coupling, erban2020coarse}, as it does not capture the
 strong temporal correlations in random forces observed from all-atom molecular
 dynamics simulations. To account for such memory effects, stochastic
 coarse-grained (SCG) models \cite{erban2016coupling, erban2020coarse} can be
 used, where auxiliary variables are introduced into the Langevin dynamics to
 model the correlated forces. Extending our current framework to incorporate
 such generalized Langevin dynamics is a natural and promising direction for
 future work, particularly for reactive systems in complex media.}

We expect that a similar analysis carries over for irreversible
reactions, e.g., $\sA+\sB\rightarrow \sC$ and $\sA+\sB\rightarrow \sC+\sD$, and
for more general networks of \rev{arbitrary} order reactions. \rev{It
would be interesting to try to adapt the work in~\cite{plesa2023stochastic}
to the reactive LD setting, investigating the commutativity of the overdamped
limit with the limit of taking the reactive interaction rates to zero within
reversible reactions (as used in~\cite{plesa2023stochastic} to approximate
higher-order irreversible reactions by systems of lower-order reversible
reactions in non-spatial models).}

\rev{Finally, we note that our asymptotic analysis was mainly focused on finite
 timescales and the interior of the domain. We did not make substantive use of
 the spatial boundary condition in the calculations (beyond assuming for the
 detailed balance calculations that the long-time spatial distribution is
 uniform). It would be an interesting future direction to study how the boundary
 conditions impact the convergence rate in $\beta$, perhaps via the formation of
 boundary layers near the domain boundary. Likewise, it is an open problem to
 fully characterize the long time behavior and its interaction with the
 overdamped limit in such reactive systems. }

\section{Code Availability}
\rev{The code used to generate the numerical results in this paper is publicly available in GitHub
at \href{https://github.com/chenyaomath/ReactLD.jl}{https://github.com/chenyaomath/ReactLD.jl}
and is archived on Zenodo
at \href{https://doi.org/10.5281/zenodo.15531388}{https://doi.org/10.5281/zenodo.15531388}.}

\section{Acknowledgments}
\rev{We thank all three anonymous reviewers for their valuable feedback that greatly improved our paper.}

\bibliographystyle{siamplain}
\bibliography{references}


\end{document}